\documentclass[%
 reprint,
 amsmath,amssymb,
 aps,
]{revtex4-1}

\usepackage{graphicx}
\usepackage{dcolumn}
\usepackage{bm}
\usepackage{color}
\usepackage{subfigure}
\usepackage{lipsum}

\graphicspath{{../}}

\begin{document}

\preprint{APS/123-QED}

\title{Nested partitions from hierarchical clustering statistical validation}

\author{Christian Bongiorno$^{(1)}$}
 \email{christian.bongiorno@centralesupelec.fr}
\author{Salvatore Miccich\`e$^{(2)}$}
\author{Rosario N. Mantegna$^{(2,3,4)}$}

\affiliation{$^{(1)}$ Laboratoire de Math\'ematiques et Informatique pour les Syst\`emes Complexes, CentraleSup\'elec, Universit\'e Paris Saclay, 3 rue Joliot-Curie, 91192, Gif-sur-Yvette, France\\
$^{(2)}$Dipartimento di Fisica e Chimica, Universit\`a di Palermo, Viale delle Scienze, Ed. 18, I-90128, Palermo, Italy \\
$^{(3)}$Complexity Science Hub Vienna, Josefst\"adter Strasse 39, 1080, Vienna, Austria \\
$^{(4)}$Computer Science Department, University College London, 66 Gower Street, WC1E 6BT, London, UK}%

\date{\today}


\begin{abstract}
We develop a greedy algorithm that is fast and scalable in the detection of a nested partition extracted from a dendrogram obtained from hierarchical clustering of a multivariate series. Our algorithm provides a $p$-value for each clade observed in the hierarchical tree. The $p$-value is obtained by computing a number of bootstrap replicas of the dissimilarity matrix and by performing a statistical test on each difference between the dissimilarity associated with a given clade and the dissimilarity of the clade of its parent node. We prove the efficacy of our algorithm with a set of benchmarks generated by using a hierarchical factor model. We compare the results obtained by our algorithm with those of Pvclust. Pvclust is a widely used algorithm developed with a global approach originally motivated by phylogenetic studies. In our numerical experiments we focus on the role of multiple hypothesis test correction and on the robustness of the algorithms to inaccuracy and errors of datasets. We also apply our algorithm to a reference empirical dataset. We verify that our algorithm is much faster than Pvclust algorithm and has a better scalability both in the number of elements and in the number of records of the investigated multivariate set. Our algorithm provides a hierarchically nested partition in much shorter time than currently widely used algorithms allowing to perform a statistically validated cluster analysis detection in very large systems. 
\end{abstract}

\flushbottom
\maketitle
%
%
\thispagestyle{empty}

\section*{Introduction}

Hierarchical clustering (HC) is a popular data analysis procedure grouping elements of a set into a hierarchy of clusters \cite{han2011data}. It is widely used in many research fields. Examples are computational biology \cite{felsenstein1981evolutionary}, genomics \cite{eisen1998cluster}, neuroscience \cite{filzmoser1999hierarchical,goutte1999clustering,baune1999dynamical}, psychology \cite{edelbrock1979mixture}, finance \cite{mantegna1999hierarchical,tumminello2007hierarchically} and economics \cite{gligor2008convergence}.  Once a dissimilarity (or similarity) measure between elements is defined and a clustering procedure is selected the hierarchical clustering algorithm is fully defined.
The algorithm is deterministic and it is providing as an output a hierarchical tree (also called dendrogram). However, the detection of a dendrogram does not mean that one also obtains a hierarchical nested partition as an output of the HC. Historically, the simplest and most popular way to obtain a partition from a hierarchical tree was to cut the dendrogram at a fixed dissimilarity value. With this simple approach, such a cut is defining the composition of clusters. They are selected by considering the groups of elements linked in the tree at a dissimilarity value smaller than the threshold value. Several methods have been proposed to select an optimal dissimilarity threshold, as the one discussed in Ref.~\cite{calinski1974dendrite}. Other authors have proposed to determine the most appropriate partition of elements by obtaining its number of clusters with different approaches. Examples are methods based on the gap statistics~\cite{tibshirani2001estimating}, squared error~\cite{jung2003decision},   connectivity~\cite{handl2005computational},  Dunn index~\cite{dunn1974well}, or silhouette width~\cite{rousseeuw1987silhouettes}. The R package clValid allows to compute hard partitions (i.e. partitions where an element can belong only to a single cluster) with  most of the previously cited methods ~\cite{brock2011clvalid}.
 The dynamical cut tree method provides a different approach, which allows a cut of the dendrogram at different distances levels~\cite{langfelder2007defining}.

Felsenstein focused on the problem of assessing the statistical significance of the clusters obtained by HC ~\cite{felsenstein1985confidence} within phylogenetic studies. Specifically, he proposed to associate a $p$-value to each clade of hierarchical tree. In phylogeny such a $p$-value provides a direct information on the evolutionary  hypothesis associated with the formation of the clade. The method used to estimate the $p$-value was based on a bootstrapping  procedure. Since the introduction of the original statistical procedure, a long debate has been ongoing in the statistical literature. Efron proposed a way to refine the test ~\cite{efron1996bootstrap}. More recently, Shimodaira implemented the refinement of Efron~\cite{efron1996bootstrap}, developed the so-called approximately unbiased (AU) test based on bootstrap~\cite{shimodaira2000another,shimodaira2004approximately}, and achieved a higher accuracy with respect to the previous proposed statistical tests. An R-package with the implementation of this test (AU test), named Pvclust, was released in Ref. \cite{suzuki2006Pvclust} and it is currently widely used in phylogenetic and genomic analyses.

It is worth noting that Felsenstein's approach is a global approach assessing the statistical reliability of the presence of all clades (i.e. groups of elements differentiating above a given value of dissimilarity) in bootstrap replicas of the original data. For this reason, the method is quite slow for large system. For some large systems the time could be so long that its application is unfeasible. Another problem of applicability of Pvclust to large set of data concerns the aspects of multiple hypothesis test correction
\cite{miller1981simultaneous}. In fact, by repeating many times a statistical test to assess the statistical reliability of the observation of each clade one needs multiple hypothesis test correction. However, the currently available multiple hypothesis test corrections, such as for example the control of the false discovery rate (FDR) \cite{benjamini1995controlling}, guarantee a highly controlled number of false positive at the expenses of a large number of false negative. Due to this limitation, the above cited Pvclust algorithm is often used without multiple hypothesis test correction opening the way to the potential presence of a number of false positive.

In addition to the methods motivated by a phylogenetic approach, other methods have been proposed more recently to associate a statistical significance to hierarchical partitions obtained by using hierarchical clustering. The primary interest for estimating such $p$-values originates in microarray expression studies but the methods proposed can be applied to any system investigated by hierarchical clustering. Examples of these approaches are the permutation test that quantifies the significance of each division of a hierarchical tree as proposed in Ref.  \cite{park2009permutation} and the comparison of similarity measures with permutation based distribution of similarity between elements obtained under the null hypothesis of no cluster in the data \cite{sebastiani2016detection}.   

In this work, we propose a greedy algorithm based on bootstrap resampling that associates a $p$-value at each clade of a hierarchical tree. Our algorithm gives good results when applied to  benchmarks mimicking the complexity of hierarchically nested complex systems \cite{mantegna1999hierarchical,tumminello2007hierarchically}.  We call our algorithm statistically validated hierarchical clustering (SVHC). Specifically, for each pair of parent and children nodes in the hierarchical tree, we test the difference between the proximity measure (in our approach a dissimilarity) associated with a clade $h$ and the dissimilarity measure associated with the clade defined by its parent node in the genealogy of the dendrogram. The statistical test we perform consider as a null hypothesis that the dissimilarity of the parent node is larger than the dissimilarity of the children node. Our tests are performed by considering multiple hypothesis test correction. In fact, we always apply the control of FDR~\cite{benjamini1995controlling}. By selecting those clades that reject our null hypothesis, we identify a hierarchically  nested partition involving a certain number of elements of the investigated systems. 
In order to evaluate the performance of our method, we test it with some benchmarks obtained by using a hierarchical factor model~\cite{schmid1957development}. In our tests, we compare our results with the ones obtained with Pvclust with and without a multiple comparison correction. Finally, we apply our algorithm to an empirical dataset. This dataset was originally obtained in Ref.~\cite{garber2001diversity} and was used as an example in the paper describing Pvclust \cite{suzuki2006Pvclust}. 

Our algorithm is highly accurate when applied to benchmarks obtained from hierarchical factor models and is also highly informative in the analysis of empirical datasets. Being our approach heuristic and local the algorithm cannot guarantee detection of global optimal solutions. This is of course a limitation of our algorithm. The positive aspect of this limitation is that our algorithm is very fast and highly scalable and therefore can be used for large datasets that would otherwise need extremely long computer time to provide results. With our algorithm one can perform a screening of large data sets, analyze results and then apply most demanding algorithms only to those sets of data that provides interesting results with a greedy approach.    


\section*{Methods}

\subsection*{Statistically Validated Hierarchical Clustering}\label{sec:test}

Let us assume that a clade originating from node $h$ has associated a dissimilarity measure $\rho_{pq}$. This is the dissimilarity value where the $p$ and $q$ children clades join in $h$. In the next step of the agglomerative algorithm, the clade originating at $h$ node joins the clade originating at  $k$ node and form the clade $l$. The dissimilarity value defining the clade $l$ is $\rho_{hk}$. The agglomerative procedure of the hierarchical clustering requires that the dissimilarity $\rho_{pq}$ must be lower than $\rho_{hk}$. This is the basic aspect of the hierarchical clustering procedure that we put at the core of our algorithm. In fact in our algorithm, for each pair of parent children clades, we perform a statistical test of the null hypothesis  $\rho_{hk} \leq \rho_{pq}$. When our null hypothesis is rejected, we consider that clade $h$ is statistically distinct from clade $l$. The $p$-value associated with each test can therefore be used to build up a nested partition where elements of statistically validated clades are elements of clusters of the partition. It should be noted that such a partition is in general a hierarchically nested partition where an element can be member of several nested clusters.
We will show below that this $p$-value can be computed analytically for Gaussian multivariate variables and numerically by computing bootstrap replicas of the dissimilarity matrix of the original data.

We consider a multivariate dataset $X$ of dimension $N \times M$ with $N$ elements and $M$ records or attributes. We call  $R$ the $N \times N$ Pearson's correlation matrix of $X$ and we use it as a similarity measure.  It is worth noting that our choice is just a possible choice of a similarity measure. In fact our procedure works for a generic definition of similarity matrix. We label as  $\sigma(x)$ the set of elements of clade defined by node $x$ of the dendrogram and $N_x$ the number of elements that clade $x$ contains. 
Hierarchical clustering is performed by using a dissimilarity measure. In this work we quantify the dissimilarity measure according to the definition 
\begin{equation}
\rho_{hk}  = \frac{ \sum_{i \in \sigma(h)} \sum_{j \in \sigma(k)} 1 - R_{ij} }{N_h\, N_k}
\label{eq:a}
\end{equation}
where $\sigma(h)$ and $\sigma(k)$ are the sets of nodes of clade $h$ and of clade $k$ in the hierarchical tree, respectively.

\subsubsection*{Analytical Derivation of the $p$-value}
We derive an analytical expression of the $p$-value $\pi_h$ associated with the null hypothesis $W_h =\rho_{hk} - \rho_{pq} \leq 0$ under the hypothesis that $X$ is a set of multivariate normal distributed random variables and $M$ is a large number. Our analytical results are obtained under the assumption that the hierarchical clustering procedure is the average linkage. Our $p$-value is defined as the cumulative distribution function in zero of the stochastic variable $W^{(s)}_h = \rho^{(s)}_{hk} - \rho^{(s)}_{pq}$, where, $\rho^{(s)}_{pq}$ is the sample mean of $\rho_{pq}$ defined as in Eq. \eqref{eq:a}. To obtain the analytical distribution of $W_h$ we notice that the distribution of a Pearson's correlation coefficient can be well approximated by a normal distribution for large values of $M$ under the assumption of normal variables.
Under all the above cited assumptions, $W_h$ is the result of a weighted sum of normal random variables. Due to the central limit theorem, the probability distribution of $W_h$ converges in probability to a normal distribution too. Since the elements of a correlation matrix are not independent variables, such sum will be a weighted sum of correlated normal random variables. In particular, according to Ref.~\cite{steiger1980tests}, the covariance between two elements $R_{i j}$ and $R_{l m}$ of a correlation matrix is 
\begin{eqnarray}
\Xi_{(i,j),(l m)} = \frac{1}{2 M} \{
\left[\left(R_{i l}-R_{i j} R_{l j}\right) \left(R_{j m}-R_{j l} R_{l m}\right)\right]+  \nonumber \\
\left[\left(R_{i m}-R_{i l} R_{l m}\right)\left(R_{j l}-R_{j i} R_{i l}\right)\right]+  \nonumber \\
\left[\left(R_{i l}-R_{i m} R_{m l}\right) \left(R_{j m}-R_{j i} R_{i m}\right)\right]+ \nonumber \\
\left[\left(R_{i m}-R_{i j} R_{j m}\right) \left(R_{j l}-R_{j m} R_{m l}\right)\right] \} \nonumber \\
\end{eqnarray}
and therefore the variance of element $R_{i j}$ is
\begin{equation}
\Xi_{(i j),(i j)} = \frac{1}{M}\left(1-R_{i j}^2\right)^2 .
\end{equation}

The expected value of the stochastic variable $W_h$ is $E[W_h] = \rho_{pq} - \rho_{hk}$. To estimate the variance of $W_h$ we must consider the covariance among the elements of the correlation matrix. Let us notice that the elements of the correlation matrix that are used to compute the average distance $\rho_{pq}$ are identified by the rectangular matrix of  $N_p$ and $N_q$ elements of sets $\sigma(p)$ and $\sigma(q)$ respectively. Similarly the elements needed to compute $\rho_{hk}$ are identified by a rectangular matrix  of elements of sets $\sigma(h)$ and $\sigma(k)$ ($N_h\,N_k$ elements). By considering the definition of  $W_h$, its variance is
\begin{eqnarray}
S[W_h]^2 = \frac{1}{\left(N_p N_q\right)^2}\sum_{i \in \sigma(p)}\sum_{j \in \sigma(q)} \sum_{l \in \sigma(p)}\sum_{m \in \sigma(q)} \Xi_{(i j),(l m)} + \nonumber \\
 \frac{1}{\left(N_h N_k\right)^2} \sum_{i \in \sigma(h)}\sum_{j \in \sigma(k)} \sum_{l \in \sigma(h)}\sum_{m \in \sigma(k)} \Xi_{(i j),(l m)}+ \nonumber \\ - \frac{1}{N_p N_q N_h N_k}\sum_{i \in \sigma(p)}\sum_{j \in \sigma(q)} \sum_{l \in \sigma(h)}\sum_{m \in \sigma(k)} \Xi_{(i j),(l m)} \nonumber \\
\end{eqnarray}

Finally the $p$-value $\pi_h$ is given by the cumulative distribution of a normal distribution with expected value $E[W_h]$ and standard deviation $S[W_h]$
\begin{equation}
\pi_{h} = P(W_h<0) = \frac{1}{2}\left[ 1 + \mbox{erf}\left(-\frac{E[W_h]}{S[W_h]\sqrt{2}} \right)\right]
\end{equation}

\subsection*{Numerical estimation of the $p$-value}
Let us call $X^{(s)}$ a bootstrap copy of $X$ obtained from sampling with replacement of the columns of $X$ matrix. Let be $R^{(s)}$ the correlation matrix obtained from a bootstrap replica.
For each bootstrap replica it is possible to compute for each group of elements a dissimilarity. For example by considering the set of nodes $\sigma(h)$ and $\sigma(k)$ we can compute the set of dissimilarity $\left\lbrace  \rho^{(1)}_{hk},\rho^{(2)}_{hk},\ldots,\rho^{(n)}_{hk} \right\rbrace $ and by considering the set of nodes $\sigma(p)$ and $\sigma(q)$ we can compute the dissimilarities $\left\lbrace  \rho^{(1)}_{pq},\rho^{(2)}_{pq},\ldots,\rho^{(n)}_{pq} \right\rbrace $ where $n$ is the number of bootstrap replicas. It is worth noting that such dissimilarities are evaluated according to the composition of sets $\sigma(x)$ by using Eq.~(\ref{eq:a}) without computing a hierarchical tree for each bootstrap replica. The $p$-value associated to cluster $h$ is defined as
\begin{equation}
\pi_{h} = \frac{\sum_{i=1}^n \delta( \rho^{(i)}_{hk}  \leq \rho^{(i)}_{pq} )}{n}
\end{equation}
\noindent
where the operator $\delta(\cdot)$ is equal to $1$ if the inequality is true, otherwise the operator is equal to $0$. In other words, the $p$-value is the fraction of times the inequality $ \rho^{(i)}_{hk} > \rho^{(i)}_{pq} $ is not satisfied in the bootstrap replicas.

Since we are computing a $p$-value for each node of the hierarchical tree we face family wise error. In this work, to perform a multiple hypothesis test correction we use the procedure of the control of the FDR~\cite{benjamini1995controlling}.  The control of the FDR procedure is implemented as follows: the $N-2$ clades are arranged in increasing order of $p$-value, labeled as $\pi^{(1)},\dots, \pi^{N-2}$. We identify the largest integer $k_{max}$ such that $\pi^{(k)} \leq k \alpha/(N-2) $ and the clades corresponding to the first $k_{max}$ $p$-values are used to build up a nested partition of the elements. The statistical threshold $\alpha$ is the maximum proportion of false discovery allowed in our statistical test. In this work we choose $\alpha=0.05$.

It should be noticed that in our algorithm the most computational demanding procedure is the computation of bootstrap replicas. Since bootstrap replicas are independent the one from the other, the algorithm can be easily and efficiently parallelized.

\begin{figure*}[tbh]
\centering
\subfigure[\label{fig:corex_no}]{\includegraphics[width=.28\linewidth]{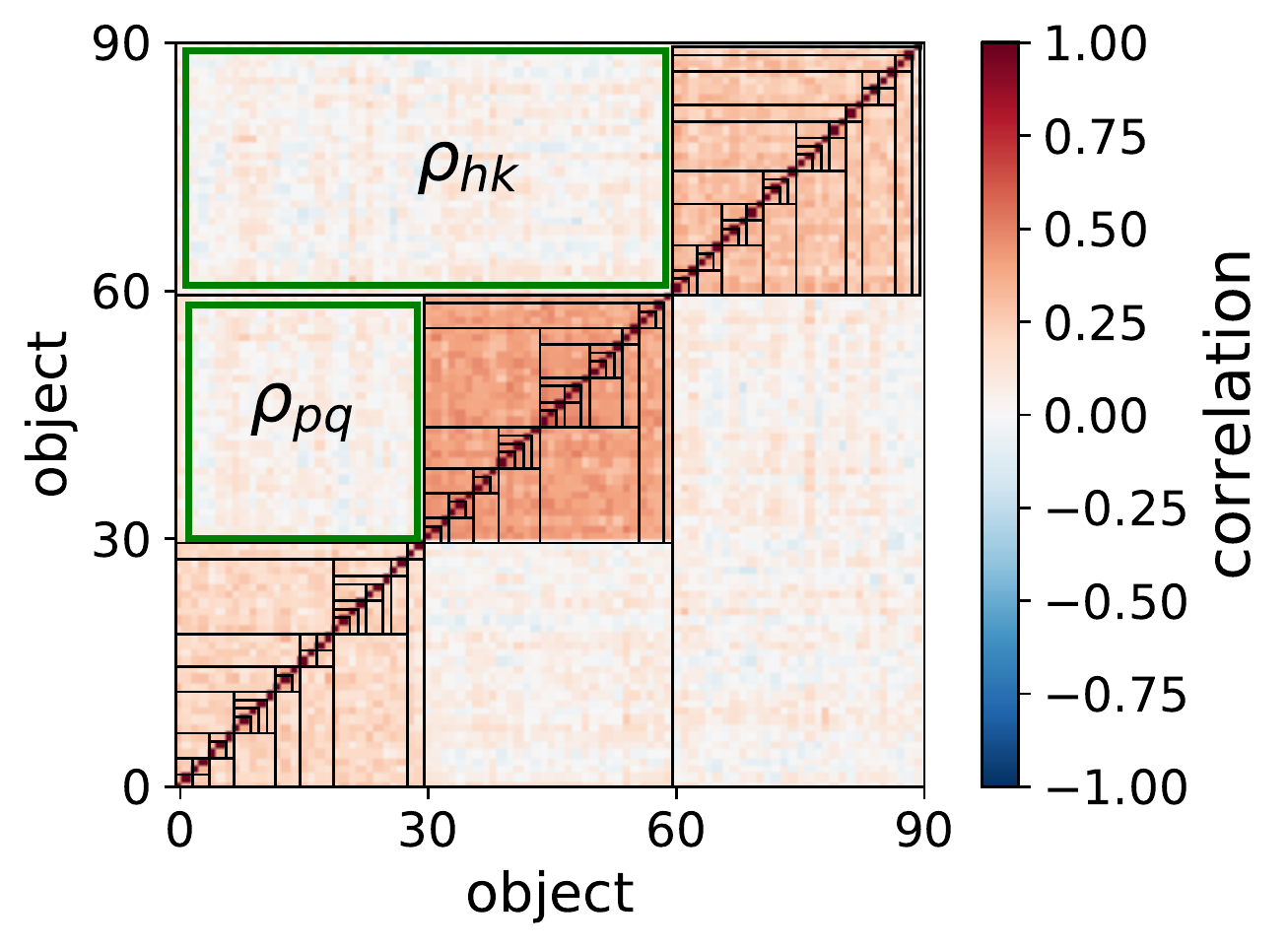} }
\subfigure[\label{fig:denr_no}]{\includegraphics[width=.28\linewidth]{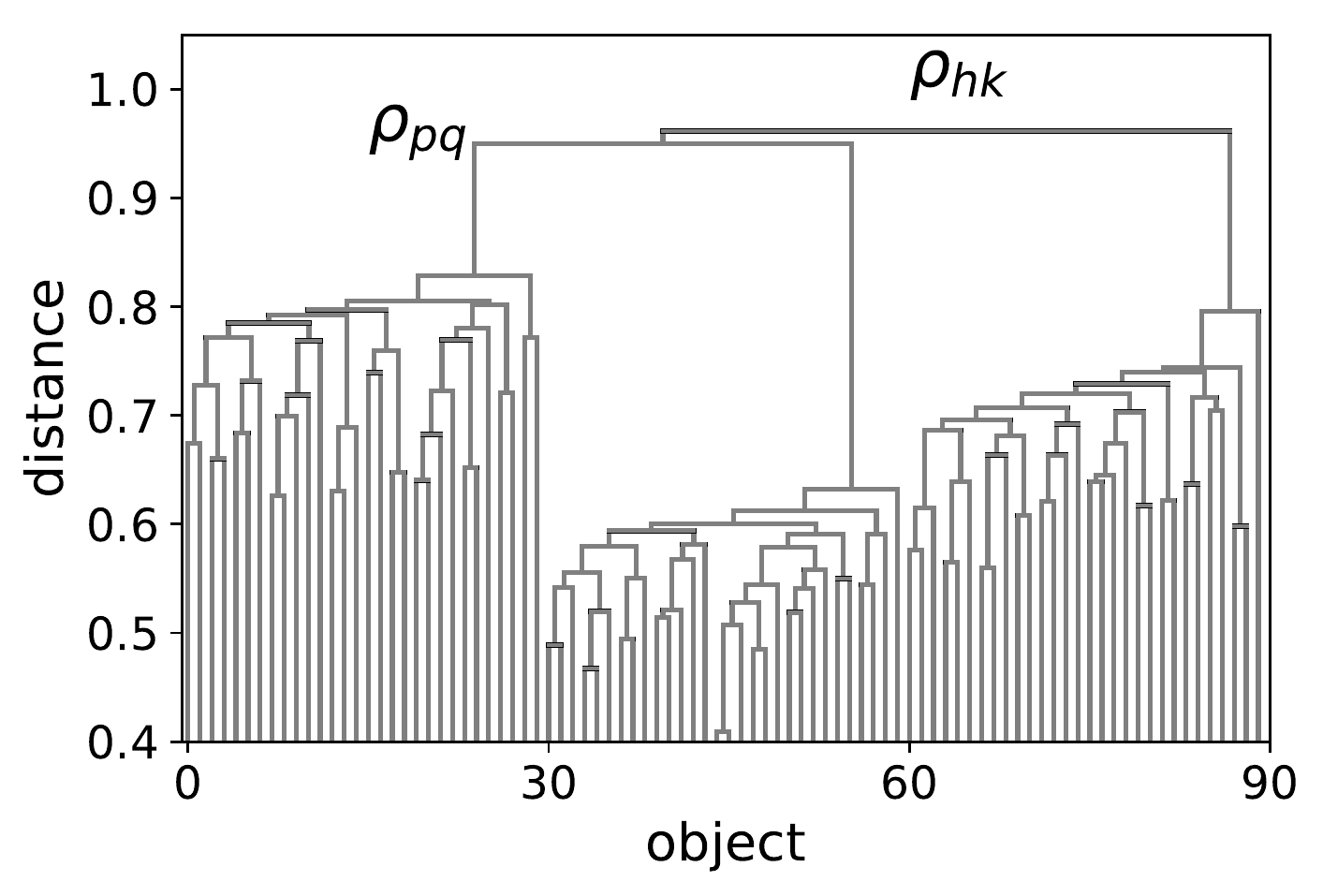} }
\subfigure[\label{fig:pv_no}]{\includegraphics[width=.28\linewidth]{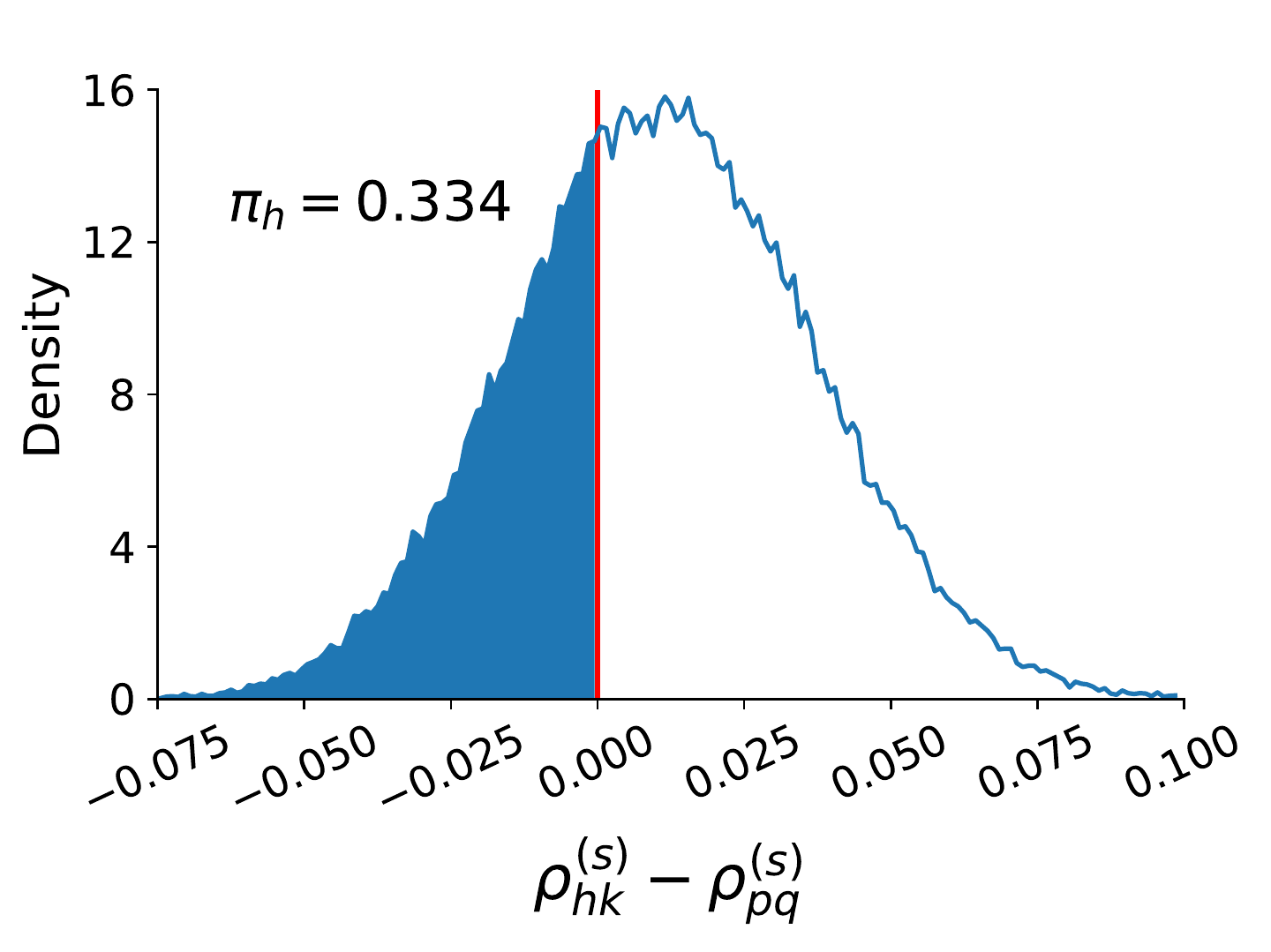} }

\subfigure[\label{fig:corex_si}]{\includegraphics[width=.28\linewidth]{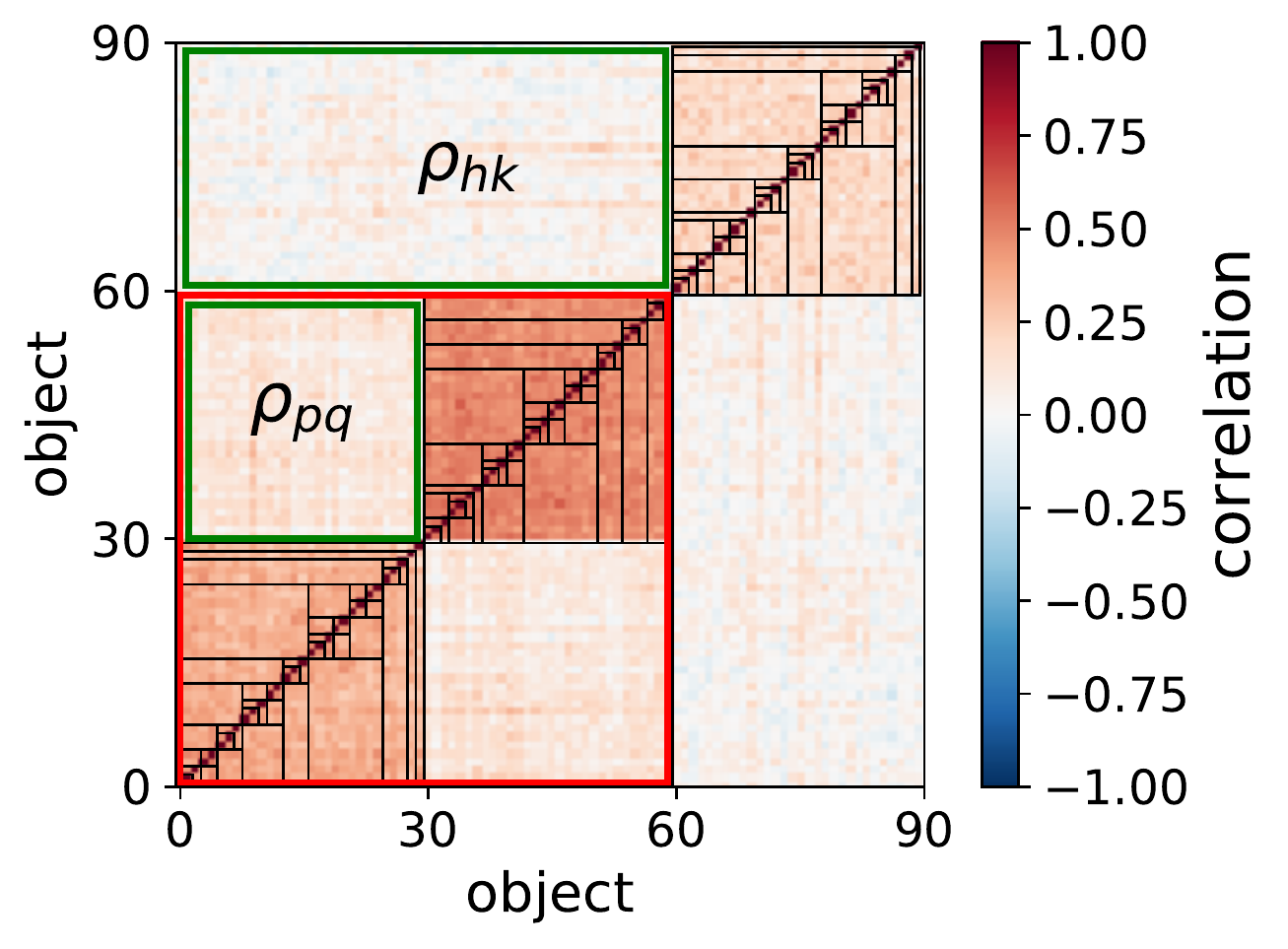} }
\subfigure[\label{fig:denr_si}]{\includegraphics[width=.28\linewidth]{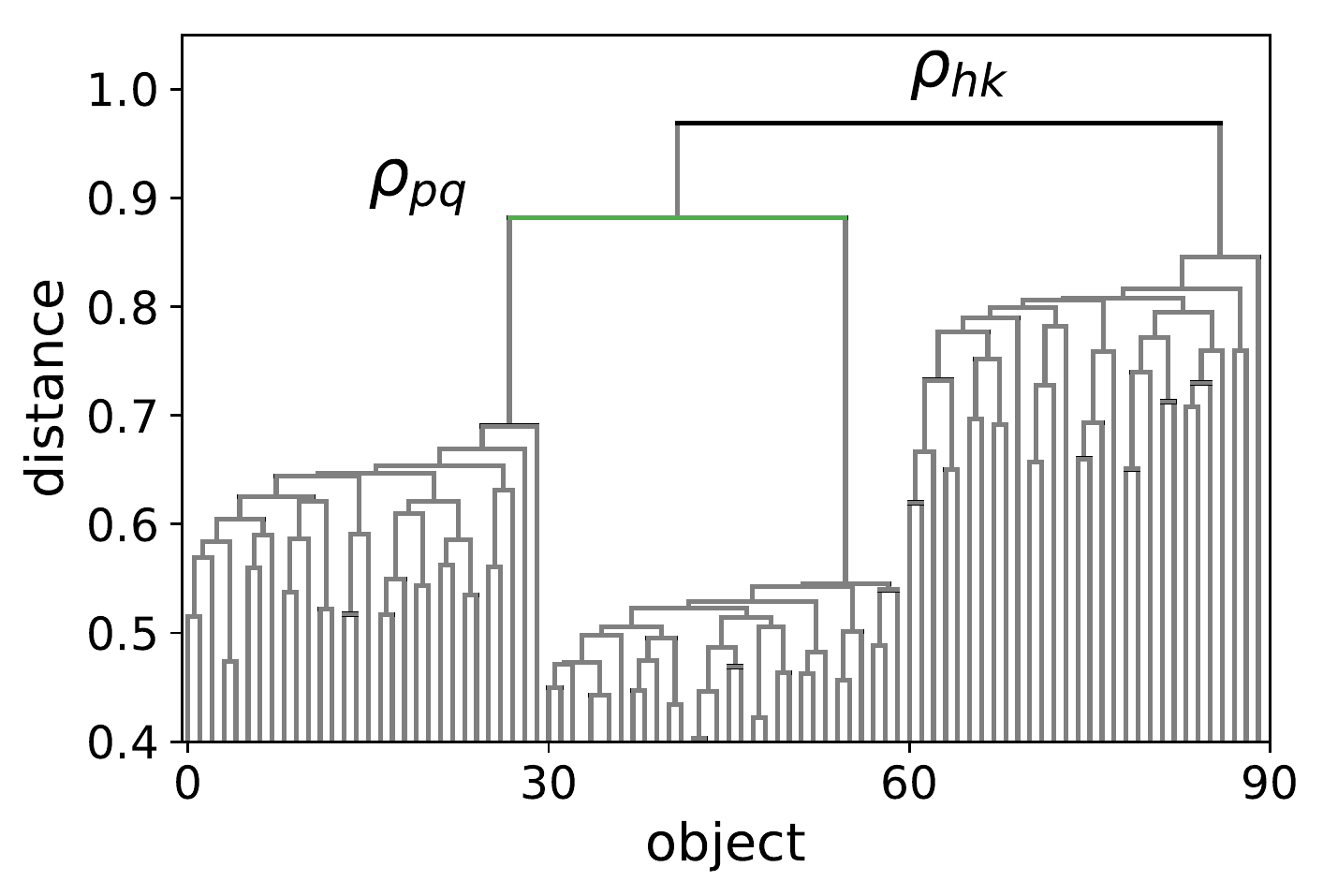} }
\subfigure[\label{fig:pv_si}]{\includegraphics[width=.28\linewidth]{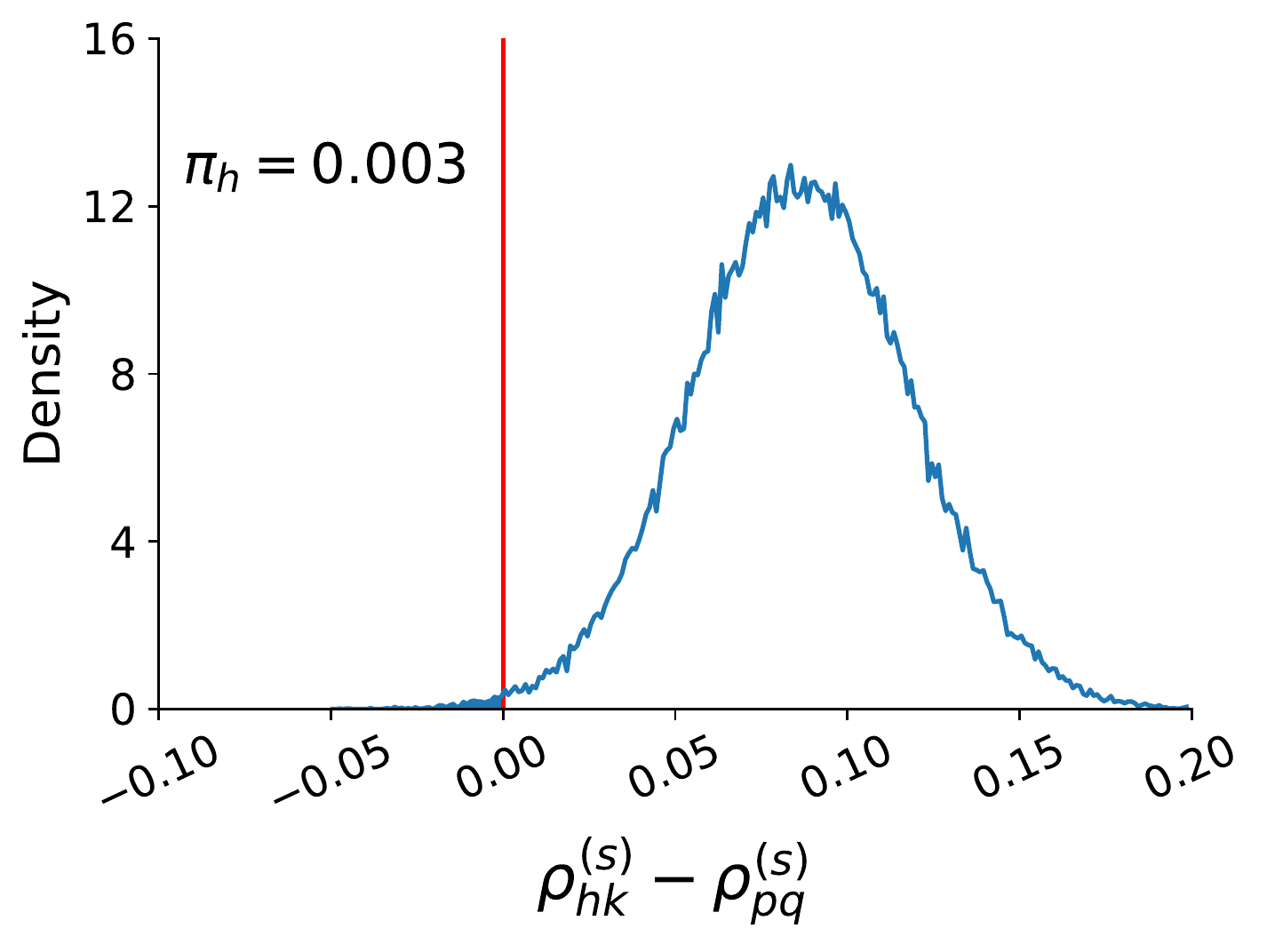} }
\caption{\small{Panels (a) and (d) show correlation matrices of slightly similar hierachically nested benchmarks generated with $M=200$. The elements are sorted according to the hierarchical tree of the HC average algorithm. The black boxes indicate the clusters of the all clades of the hierarchical tree. The green boxes highlight the correlation coefficients used to evaluate $\rho_{hk}$ and $\rho_{pq}$. The red box of panel (d) indicates that cluster of clade $h$ is statistically validated. The elements $[0,29]$ belongs to the set of elements $\sigma(p)$, the elements $[30,59]$ belongs to the set of elements $\sigma(q)$, the elements $[60,89]$ belongs to the set of elements $\sigma(k)$, and the elements $[0,59]$ belongs to the set of elements $\sigma(h)$.  In panels (b) and (e) we show dendrograms of average HC of dissimilarity matrices associated with the multivariate datasets with correlation matrices of panels (a) and (d) respectively. In panels (c) and (f)  we show the density function of $\rho_{hk}^{(s)}-\rho_{pq}^{(s)}$ to illustrate how the $p$-value of the null hypothesis  $W_h \leq 0$ is estimated (in these examples we perform $n=100,000$ bootstrap replicas). In the example of panels (a), (b), and (c)  the null hypothesis $W_h  \leq 0$ is not rejected whereas in the example of panels (d), (e), and (f) the same null hypothesis is rejected }}\label{fig:ex1}
\end{figure*}

To illustrate our procedure of numerical estimation of the $p$-value, we show two examples of statistical validation of a clade in Fig.~\ref{fig:ex1}. 
Specifically, we consider two slightly different sample hierarchical trees. They are shown in Fig.~\ref{fig:denr_no}  and Fig.~\ref{fig:denr_si} respectively. The test aims to evaluate whether the elements from 0 to 59 (i.e. the clade originating at the node of the dendrogram characterized by the $\rho_{pq}$ dissimilarity) are defining a group of stocks statistically distinct from the set of all stocks.
In the top row of Fig.~\ref{fig:ex1} we show three panels referring to the case when the null hypothesis 
$W_h  \le 0$ is not rejected and therefore the clade of elements from 0 to 59 cannot be considered as a group of elements hierarchically distinct from all elements. According to the hierarchical tree, the clade of elements  $\sigma(p)$  (elements $[0,29]$) and the clade of elements $\sigma(q)$ $[30,59]$) join together in the clade  $\sigma(h)$, originating at $\rho_{pq}=0.95$.  Then the clade $\sigma(h)$  joins with clade $\sigma(k)$  (composed by the element $[60,89]$) at the node characterized by the dissimilarity $\rho_{hk}=0.96$. In the sample tree, the dissimilarity value $\rho_{pq}=0.95$ is smaller then $\rho_{hk}=0.96$, as shown in Fig.~\ref{fig:denr_no}. However, in spite of this structure observed in the hierarchical tree of the sample correlation matrix, the bootstrap analysis of $\rho^{(s)}_{pq}$ and $\rho^{(s)}_{hk}$ shows that the null hypothesis  $W_h \le 0$ has associated a $p$-value equal to $\pi_h=0.333$ and therefore cannot be rejected (see Fig.~\ref{fig:pv_no}). For this example, we therefore conclude that the set of elements [0,59] cannot be distinguished from the set of elements [0,99]. 

In the bottom row of Fig. \ref{fig:ex1} we show a slightly different example. Specifically, in this case the dissimilarity values are $\rho_{pq}=0.88$ and $\rho_{hk}=0.96$, as shown in Fig.~\ref{fig:denr_si}. In other words, elements [0,59] are slightly more correlated than in the previous case. For this set of data, our approach concludes that the clade [0.59] is statistically distinct from the complete set [0.99] since the inequality $W_h \leq 0$ is  verified (Fig.~\ref{fig:pv_si}) only for $0.3\%$ of our bootstrap replicas. Therefore the null hypothesis  $W_h \leq 0$ has associated a $p$-value  $\pi_h=0.003$ and after performing the FDR multiple hypothesis test correction we reject it.

\subsection*{Hierachically nested Benchmark}\label{sec:hnbench}
In our numerical experiments, We use benchmarks of multivariate datasets obtained with a nested factor model with $r$ common factors~\cite{schmid1957development}. Specifically, We simulate a multivariate dataset $X$ of $N$ elements with $M$ records by using the equation
\begin{equation}
X_{ij} = \sum_{k=1}^{r}P_{ik}\,A_{kj}+U_i\,\varepsilon_{ij}
\end{equation}
where $P$ is the factor loading matrix of dimension $N \times r$ and $A$ is a factor score matrix of dimension $r \times M$ with entries that are standardized independent Gaussian variables orthogonalized with a Gram-Schmidt algorithm. The vector $U_i$ is called uniqueness and it is given by $U_i = \sqrt{1 -\sum_{j=1}^r P_{ij}^2}$. Finally, $\varepsilon_{ij}$ is also a standardized Gaussian variable.

\begin{figure}[tbh]
\centering
\subfigure[\label{fig:pattex}]{\includegraphics[width=.51\linewidth]{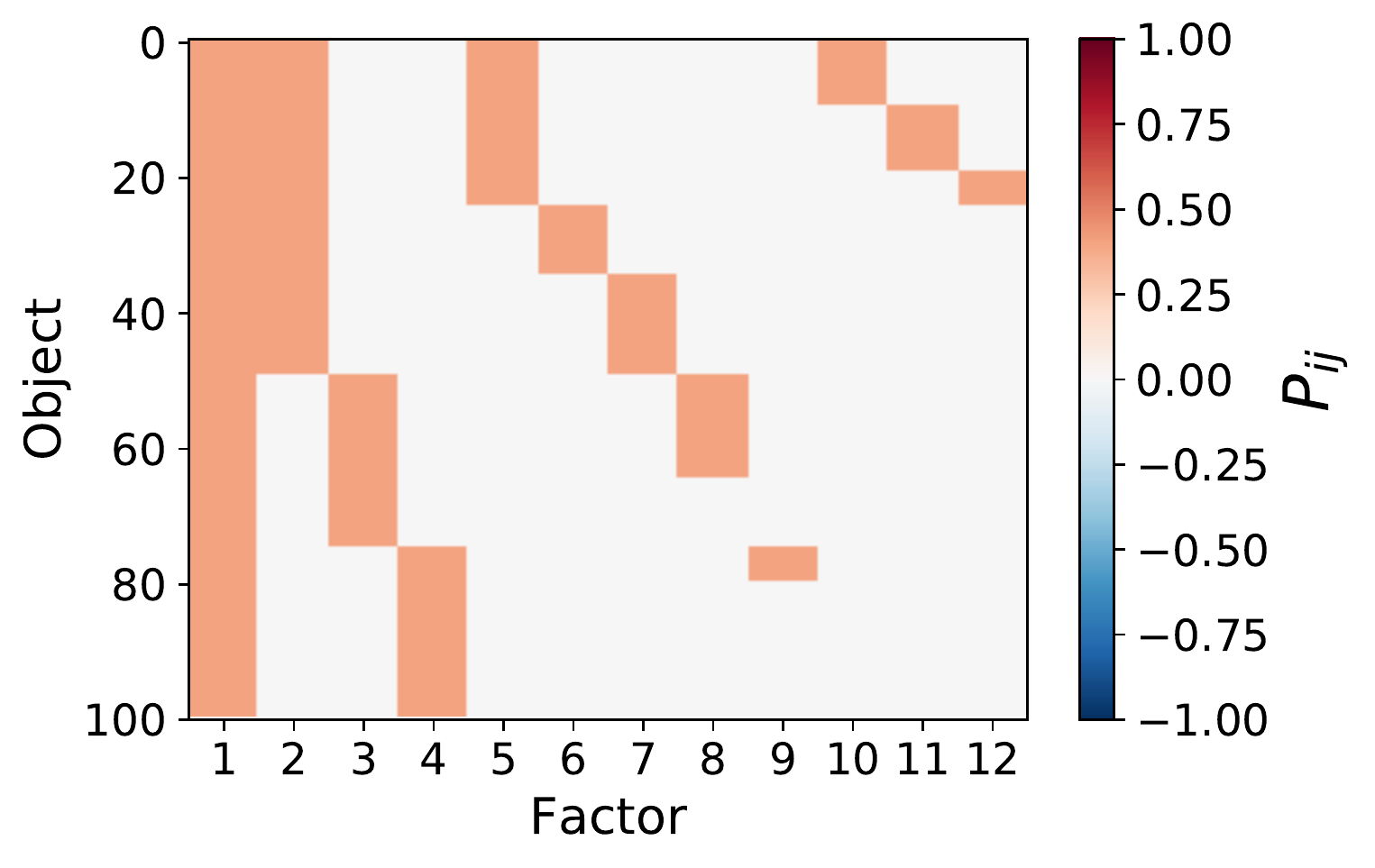} }
\subfigure[\label{fig:correx}]{\includegraphics[width=.45\linewidth]{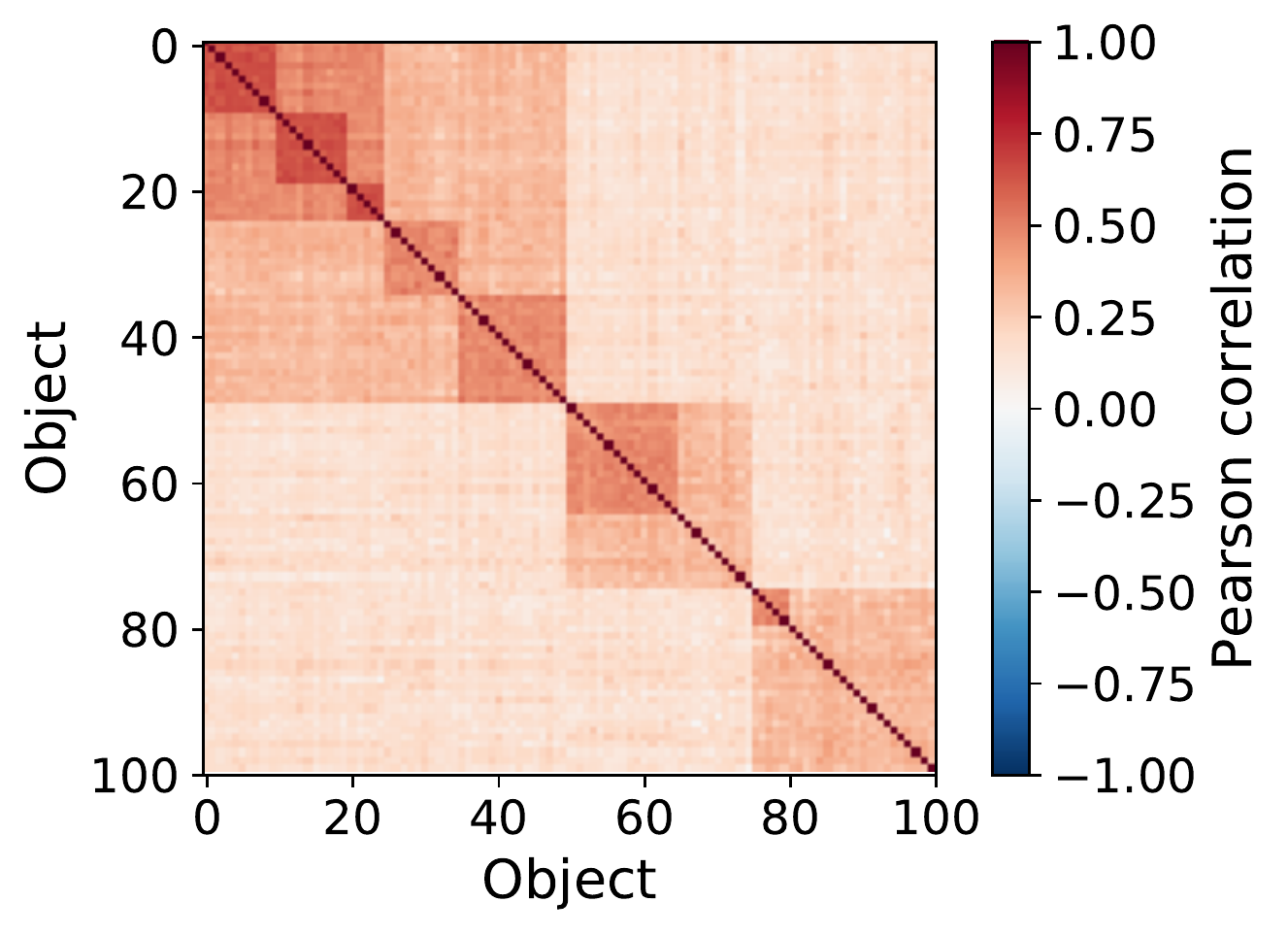} }
\caption{$(a)$ Example of factor loading pattern matrix. $(b)$ Person's correlation matrix obtained from a multivariate dataset obtained by using the factor loading matrix of (a) with with $r=12$ and a factor score matrix $r \times M$  with  $r=12$ and $M=500$ standardized independent Gaussian variables.}\label{fig:bench}
\end{figure}

A nested factor model is able to generate a multivariate set characterized by a correlation matrix showing hierarchically nested blocks. For example, the multivariate dataset $X$ obtained from the factor loading matrix $P$ of Fig.~\ref{fig:pattex} with $N=100$ elements and $12$ factors together with a factor score matrix $A$  with 12 factors and $M=500$ records has associated the correlation matrix shown in Fig.~\ref{fig:correx}.
With this choice of $P$ each factor corresponds to a block on the correlation matrix, and an element $i$ is a member of the block associated with the $r$ factor if $P_{ir}$ has a positive value. Specifically, the factor loading matrix of Fig.\ref{fig:pattex} has all positive elements equal to $0.4$, and it produces twelve blocks of sizes $\{100,\, 50, \,25, \,25, \,25, \,10, \,15,  \,15, \,  5, \, 10, \, 10, \,  5\}$. In some numerical experiments we add noise to each record to investigate the robustness of the algorithms to inaccuracy and errors of the datasets. This is done by computing 
$$X_{ij} = (1-\lambda) X'_{ij} + \lambda \varepsilon_{ij}$$
where $X'_{ij}$ is the dataset without noise, $\varepsilon$ is a standardized Gaussian variable, and $\lambda \in [0,1]$ is the parameter controlling the amount of noise inserted into the dataset.

In numerical experiments discussed in the Results section, we are using the factor loading matrix of Fig.~\ref{fig:pattex} and a number of modifications of it. However, we have tested the robustness of our results for many other  factor loading matrices.

\subsection*{Comparing partitions}\label{sec:onmi}
The comparison metric used in this paper to assess the similarity of two hierarchically nested partitions is the overlapping normalized mutual information (ONMI) ~\cite{mcdaid2011normalized}. ONMI is a variant of the normalized mutual information (NMI) \cite{danon2005comparing}. $\mbox{NMI}(x,y)$ measures the amount of information obtained about a partition $x$ through the knowledge of another partition $y$, or vice-versa. NMI was defined to compare hard partitions. It was generalized to compare overlapping partitions in Ref.~\cite{lancichinetti2009detecting}. Later authors of Ref.~\cite{mcdaid2011normalized}  proposed the modification of the ONMI metric that we are adopting in this paper. It is worth stressing that a hierarchical partition is a special case of an overlapping partition, with overlapping groups constrained to be nested.

\section*{Results}

\subsection*{Comparison between the analytical and bootstrap based $p$-value}\label{sec:comp}
We first report a numerical experiment comparing  the bootstrap based $p$-value $\pi_h^{(b)}$ of our algorithm with the analytic $p$-value $\pi_h^{(a)}$ for Gaussian and Student's t multivariate variables.
 
\subsubsection*{The Gaussian case}
\begin{figure}[tbh]
\centering
\subfigure[\label{fig:distwGaus}]{\includegraphics[width=.42\linewidth]{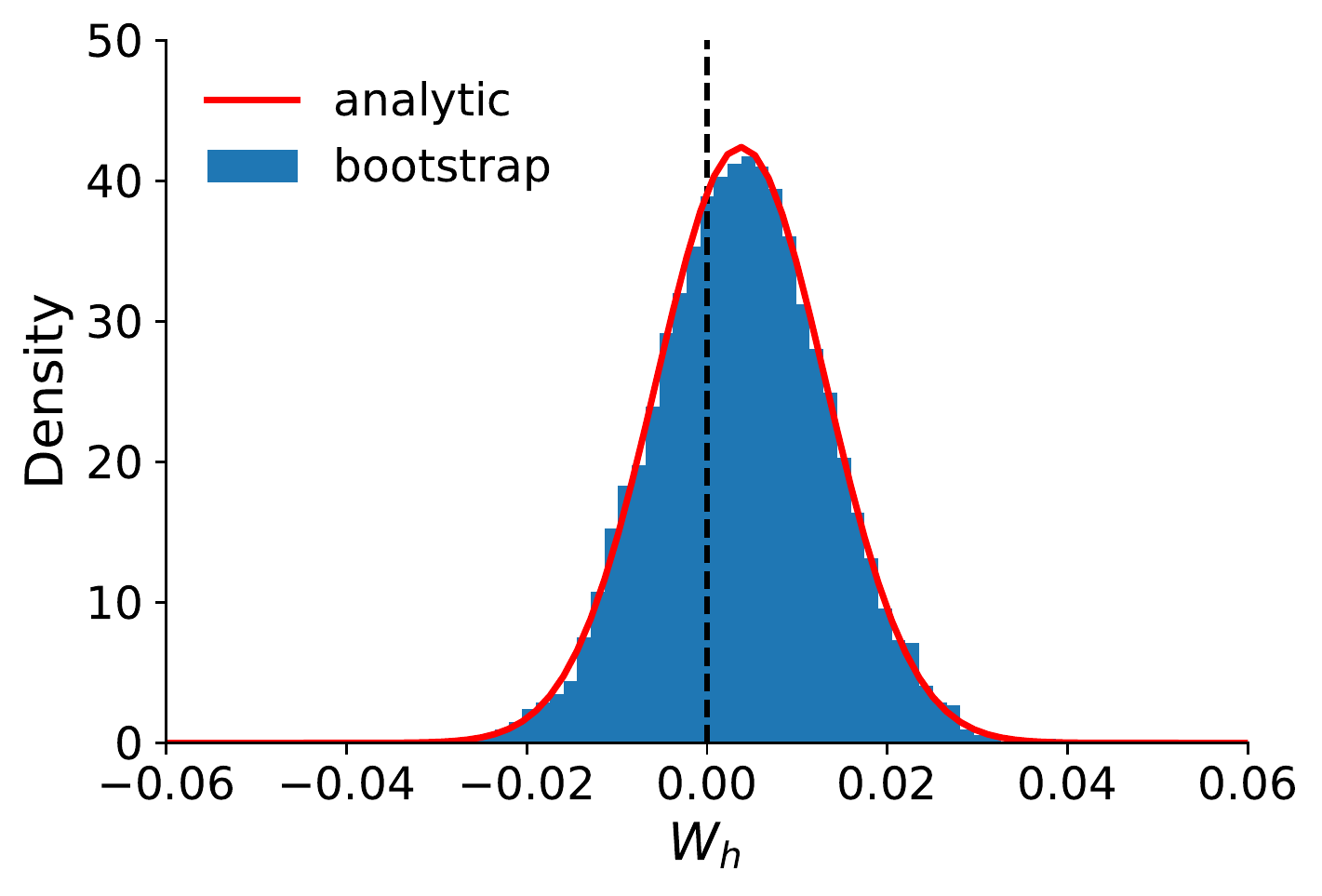} }
\subfigure[\label{fig:bvsaGaus}]{\includegraphics[width=.42\linewidth]{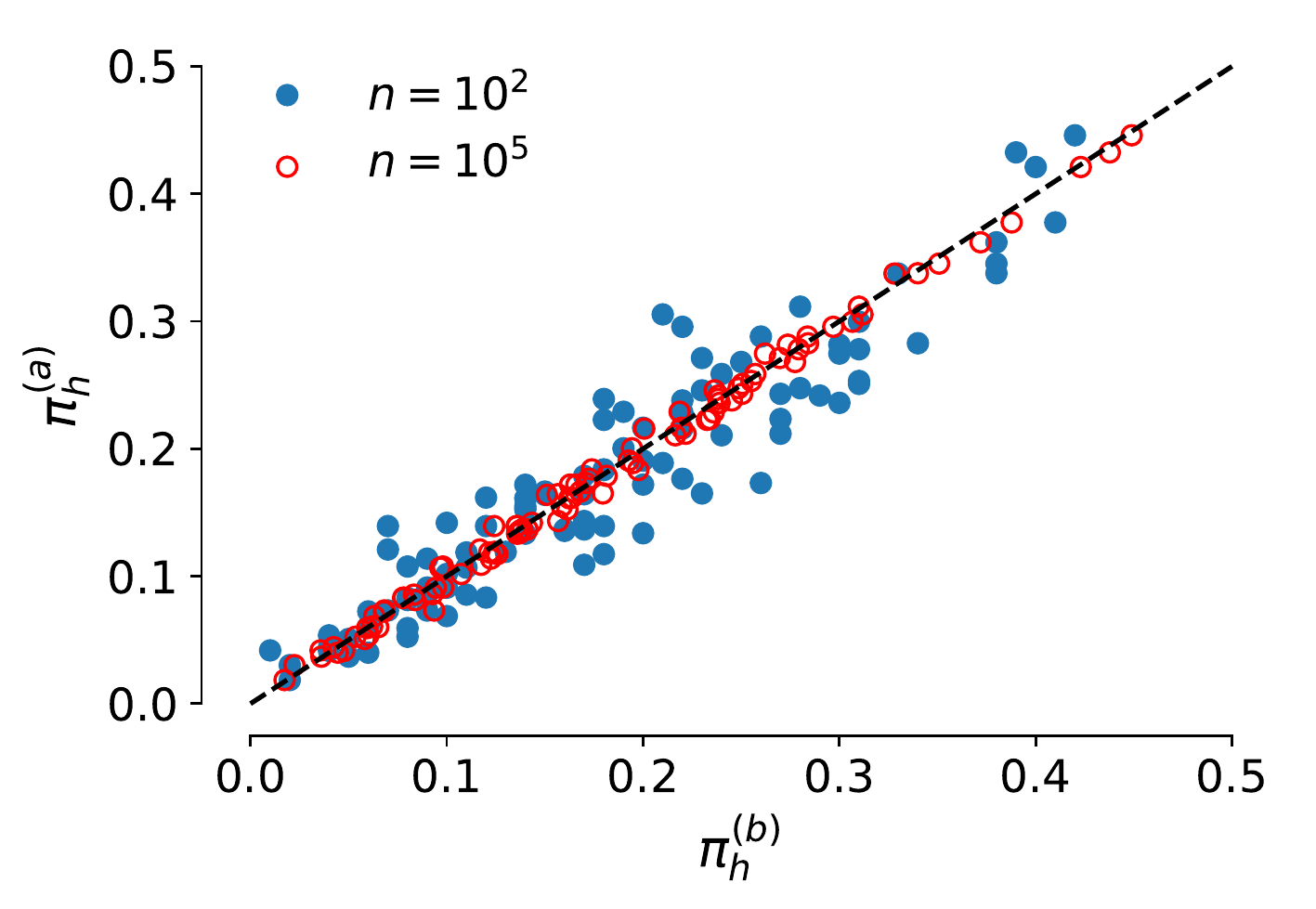} }

\subfigure[\label{fig:distwT}]{\includegraphics[width=.42\linewidth]{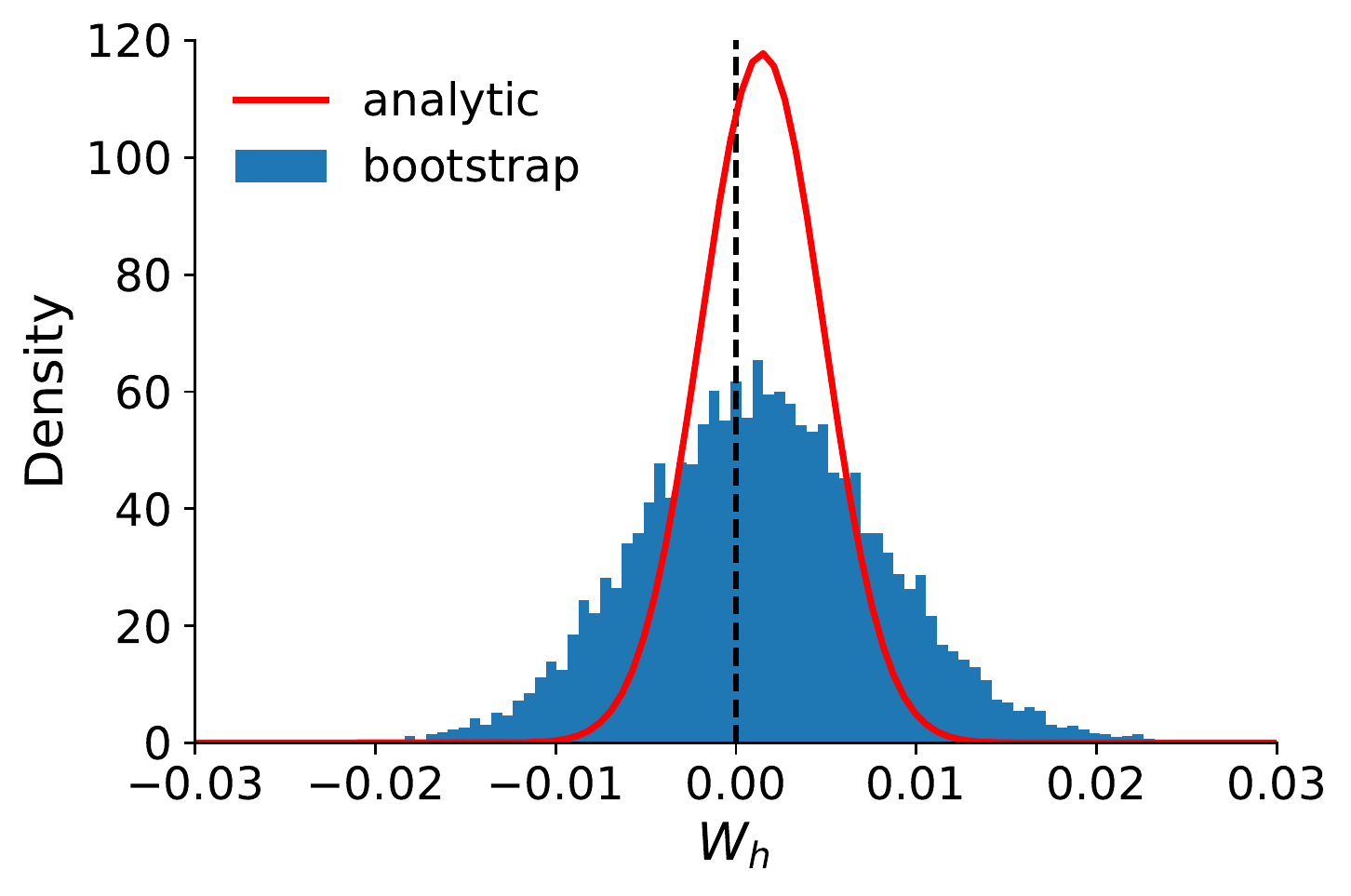} }
\subfigure[\label{fig:bvsaT}]{\includegraphics[width=.42\linewidth]{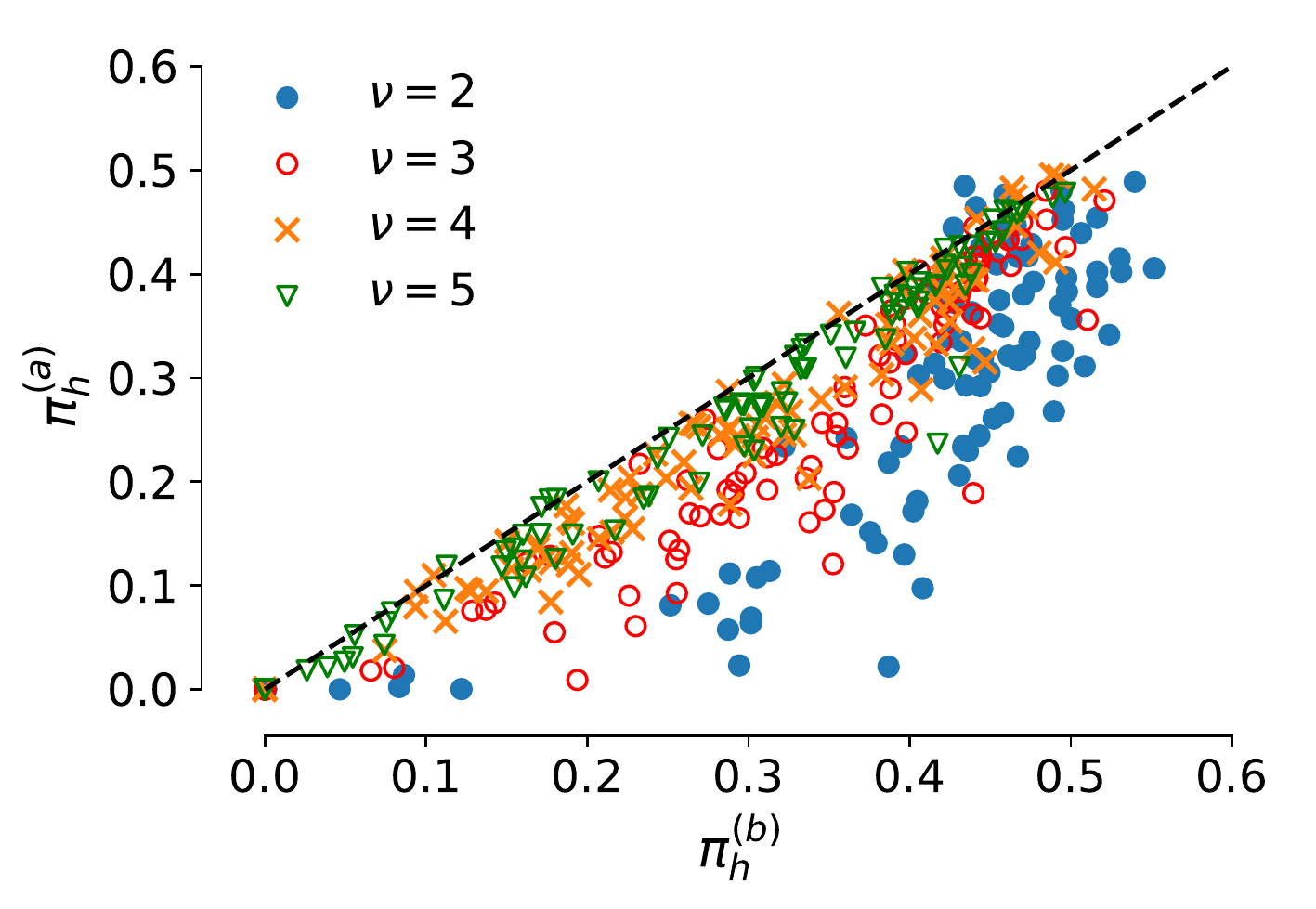} }
\caption{Numerical experiments performed with uncorrelated multivariate random variables. $(a)$  Histogram of the probability density function of $W_h$ for a selected clade obtained by using  $n=10^4$ bootstrap replicas of Gaussian multivariate random variables. The red line is the analytical probability density function obtained under the hypothesis of Gaussian variables.$(b)$ Scatter plot of e $p$-values of analytic computation $\pi_h^{(a)}$ versus $p$-values obtained with bootstrap $\pi_h^{(b)}$  for two values of $n$. Each point refers to the $p$-value of a clade (Gaussian random variables). $(c)$ Histogram of the probability density function of $W_h$ for a selected clade obtained by performing $n=10^4$ bootstrap replicas. The red line is the analytical probability density function expected for Gaussian variables for Student's t multivariate random variables. $(d)$ Scatter plot of $\pi_h^{(a)}$ versus $\pi_h^{(b)}$ for different values of the parameter $\nu$ of Student's t random variables. Each point refers to the $p$-value of a clade.}\label{fig:gauss}
\end{figure}

We numerically generate a set of multivariate uncorrelated Gaussian random variables $X$. The set has $N=100$ elements with $M=1000$ records each. Our numerical experiment is done  for different values of the number of bootstrap replicas $n$. In Fig.~\ref{fig:distwGaus}, we show an example of the bootstrap probability density function of the stochastic variable $W_h$ for a selected clade $h$ compared with the result of the analytical computation. In Fig.~\ref{fig:bvsaGaus}, we also show a scatter plot  between $\pi_h^{(a)}$ and $\pi_h^{(b)}$ of one bootstrap realization for two values of $n=(10^2,10^5)$.  It is worth noting that the bootstrap $p$-values converge to their analytical values for large values of $n$. In our numerical experiments we do not detect any bias in the numerical estimation of bootstrap $p$-values for Gaussian multivariate data.

\subsubsection*{Student's t-distribution case}
In order to study the sensitivity of the analytic estimation to the Gaussian hypothesis we compare analytical and bootstrap $p$-values in the case of a multivariate dataset  $X$  of uncorrelated t-distributed random variables of $N=100$ elements with $M=1000$ records each.
The probability density function of a $t$-distributed variable is
\begin{equation}
f(x) = \frac{\Gamma(\frac{\nu+1}{2})} {\sqrt{\nu\pi}\,\Gamma(\frac{\nu}{2})} \left(1+\frac{x^2}{\nu} \right)^{\!-\frac{\nu+1}{2}}.
\end{equation}
The parameter $\nu$ controls the finiteness of main moments. Specifically, for $1<\nu \leq 2$ the variance is not defined, for $2<\nu \leq 4$ the variance is finite, but the kurtosis is not defined and for $\nu>4$ both variance and kurtosis are finite. It is worth recalling that a value of $\nu \approx 3$ has been observed in several financial studies~\cite{gopikrishnan1998inverse}.\\

In Fig.~\ref{fig:distwT}, we show an example of the bootstrap probability density function of the stochastic variable $W_h$ for a selected clade $h$ compared with the analytic probability density function expected for Gaussian variables. We note that the analytical Gaussian $p$-value underestimates the variance of the stochastic variable $W_h$. This conclusion is confirmed by inspecting Fig.~\ref{fig:bvsaT} where we show a scatter plot between $\pi_h^{(a)}$ and $\pi_h^{(b)}$ 
for different values of $\nu$. It is worth noting that for large value of $\nu$ the discrepancy between analytical and bootstrap $p$-values  is progressively reduced since the t-distribution converges to the Gaussian  when $\nu \to \infty$~\cite{lange1989robust}. We therefore conclude that numerical investigations are therefore essential when the probability density function of the multivariate dataset differs from a Gaussian multivariate.

\subsection*{Experiments on the Benchmark}\label{sec:resbench}
Here we investigate the effectiveness of our algorithm in retrieving the true hierarchical nested partition of a representative benchmark. We also compare our results with the outputs of the algorithm Pvclust ~\cite{suzuki2006Pvclust}. 
We show that the SVHC algorithm has a good scalability for large systems. For this reason, by considering that our algorithm is preferentially suggested to investigate large systems a multiple hypothesis test correction is part of the algorithm. On the other hand, the multiple hypothesis test correction is an option in the Pvclust algorithm. To take into account this important difference, we are comparing the two algorithms by considering the SVHC output and two outputs of Pvclust, the first obtained without multiple hypothesis test correction (labeled in our figures as "single") and the second obtained with the control of the FDR (labeled as "FDR"). Partitions investigated in this paper are by construction hierarchically nested partitions. We therefore compare hierarchically nested partitions. 

\subsubsection*{Hierarchical partitions}
\begin{figure}[tbh]
\centering
\subfigure[\label{fig:noiseONMI}]{\includegraphics[width=.42\linewidth]{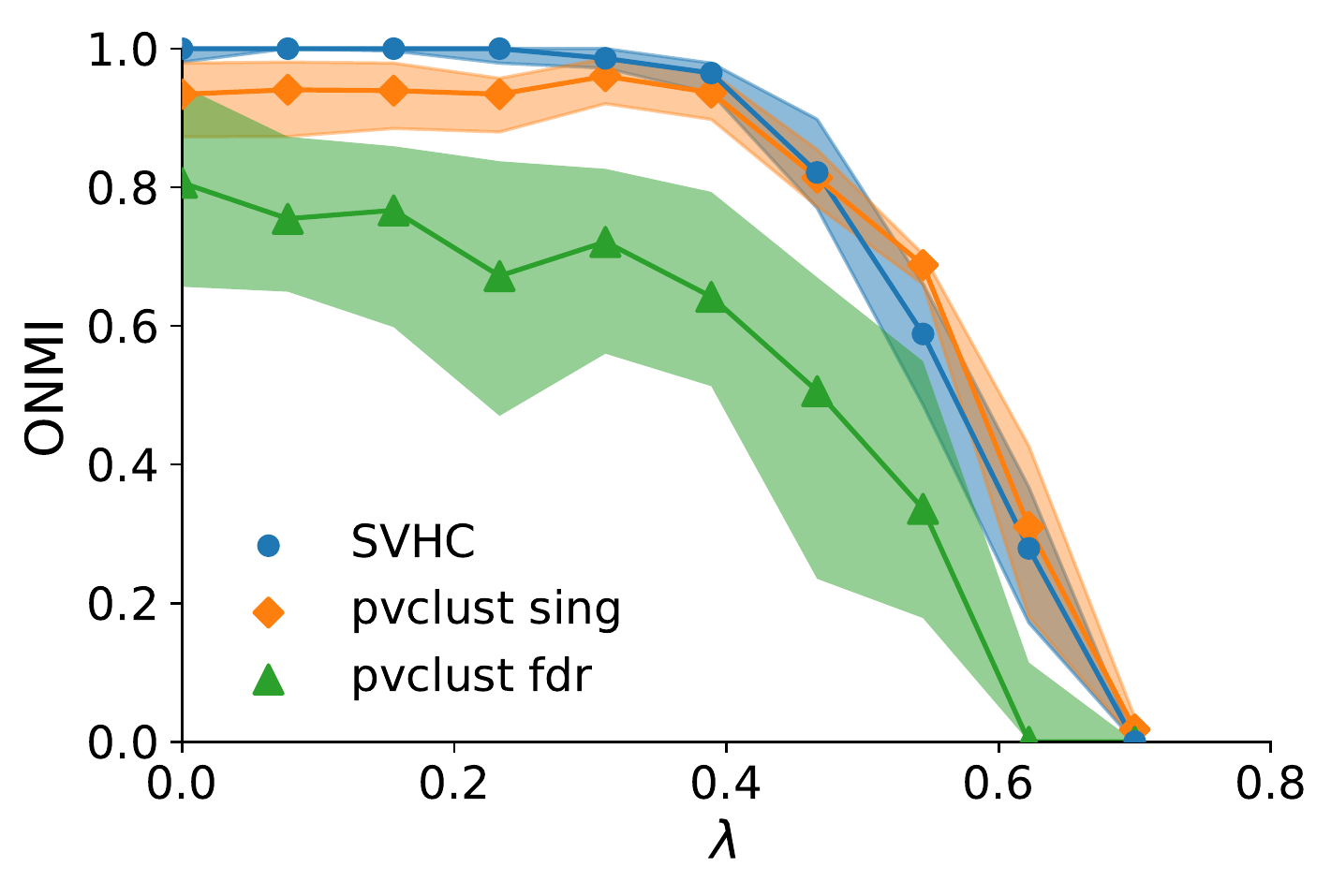} }
\subfigure[\label{fig:noiseNc}]{\includegraphics[width=.42\linewidth]{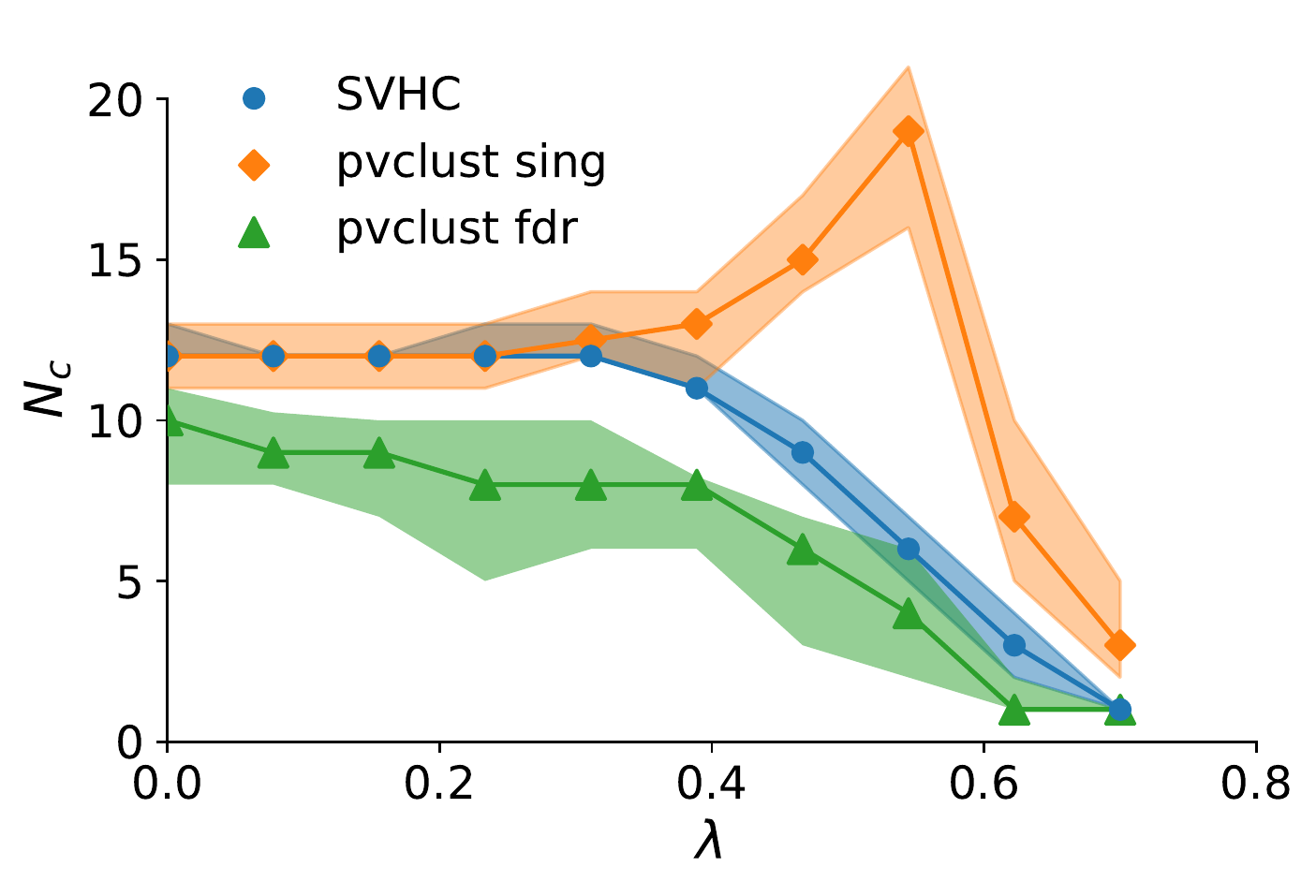} }
\caption{
$(a)$ Overlapping normalized mutual information (ONMI) between the true hierarchical partition of the benchmark of Fig.~\ref{fig:bench} with $N=100$ and $M=500$ and the hierarchical partition obtained with SVHC, Pvclust "single" or Pvclust "FDR"  as a function of the noise parameter $\lambda$. $(b)$ Number of statistically validated clusters detected by the algorithms as a function of the noise parameter $\lambda$. Points are the median computed in $100$ independent realizations. The color band highlights the interval between the $25$ and the $75$ percentile. In our numerical experiments, we simulate $1000$ bootstrap replicas both for the SVHC and the Pvclust algorithm.}\label{fig:bench6}
\end{figure}

In a second set of numerical experiments (see Fig.~\ref{fig:bench6}), we explore the robustness of the two algorithms to different levels of noise in the detection of hierarchically nested partitions of the benchmark. 
For low levels of noise ($\lambda \le 0.4$) the SVHC algorithm has a very good performance both in terms of ONMI with the true partition of the benchmark (see Fig.~\ref{fig:noiseONMI}) and in terms of the number of clusters detected (see Fig.~\ref{fig:noiseNc}). In the analysis of the figure, it should be noted that the benchmark is characterized by 12 nested clusters. Pvclust "single" has a similar performance for low values of $\lambda$ but the quality of the detected hierarchical partition is strongly affected by the option about the multiple hypothesis test correction. In fact, in Fig.~\ref{fig:noiseONMI} we  observe that the hierarchical partition of Pvclust "FDR"  has lower performance than the one of Pvclust "single".

For values of the noise parameter $\lambda >0.4$ both the SVHC and the Pvclust algorithms reduce their ability to retrieve the true hierarchical partition  and they have similar performances concerning the ONMI metric. However, the outputs obtained by the two algorithms are characterized by a different types of error. In fact, in Fig.~\ref{fig:noiseNc} we show that starting from noise parameter $\lambda \approx 0.4$ the output of the algorithm Pvclust "single" starts to be characterized by an increasing number of clusters whereas the SVHC presents the opposite case. As a result, in statistical terms the partition retrieved from the SVHC algorithm is more precise than the one obtained from Pvclust "single" , although it lacks in the recall (i.e. it has a large number of false negative). 

In a third set of numerical experiments we investigate the effectiveness of algorithms in retrieving the true hierarchical  partition as a function of the number of elements $N$ of the system. In these experiments we again use a benchmark with twelve nested clusters. This is done by using the same type of benchmark of Fig.~\ref{fig:bench} modified by increasing the number of elements of each cluster and the total number of elements proportionally. Moreover, the number of records $M$ of the time series is also increased proportional to $N$ according to  $M=5N$. We perform numerical experiments for systems of sizes equal to $N=\{56, 100, 178, 316, 562\}$ where the different values present a logarithmic spacing. In this set of experiments noise is absent ($\lambda=0$).

\begin{figure*}[tbh]
\centering
\subfigure[\label{fig:onmiNvar}]{\includegraphics[width=.32\linewidth]{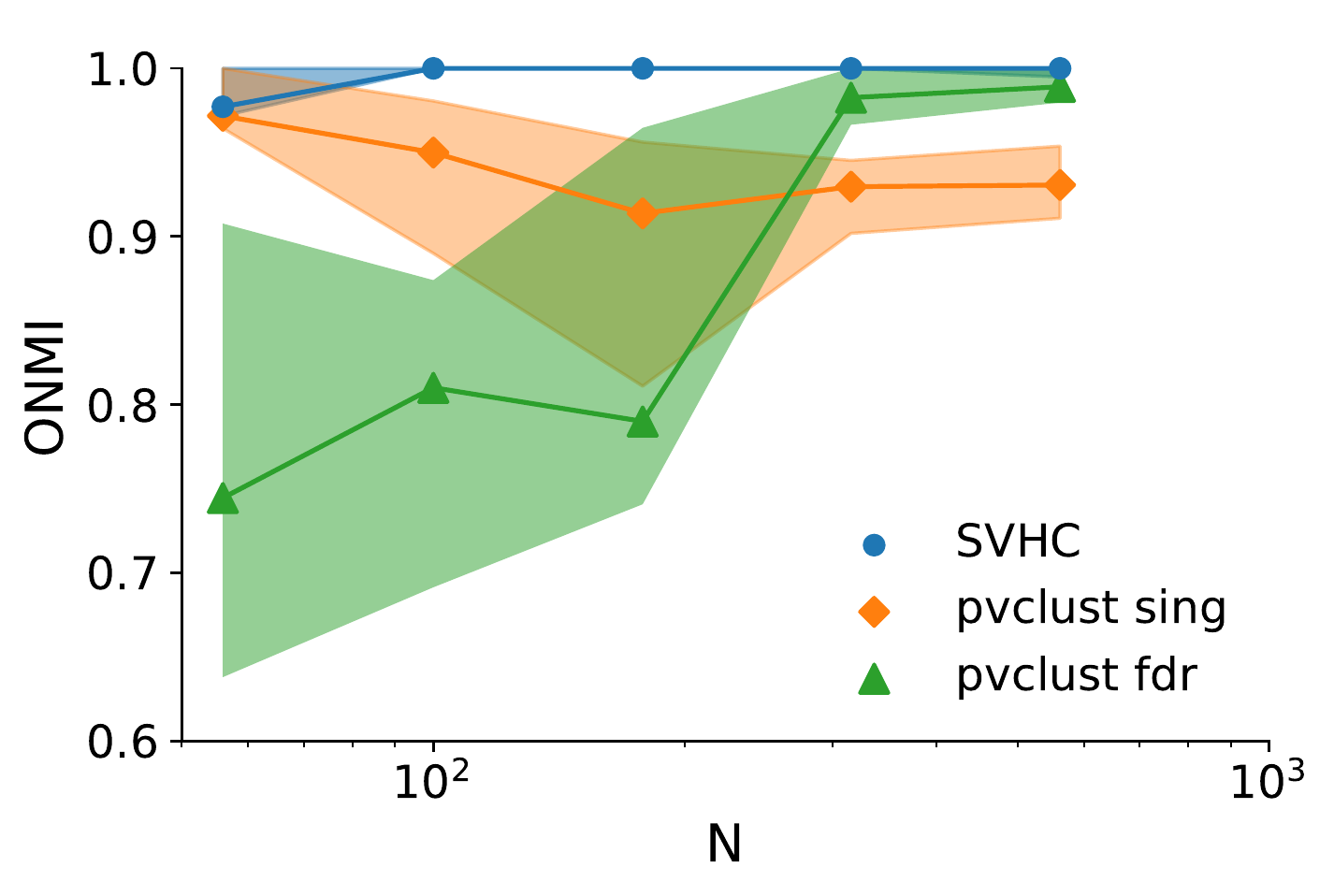} }
\subfigure[\label{fig:ncNvar}]{\includegraphics[width=.32\linewidth]{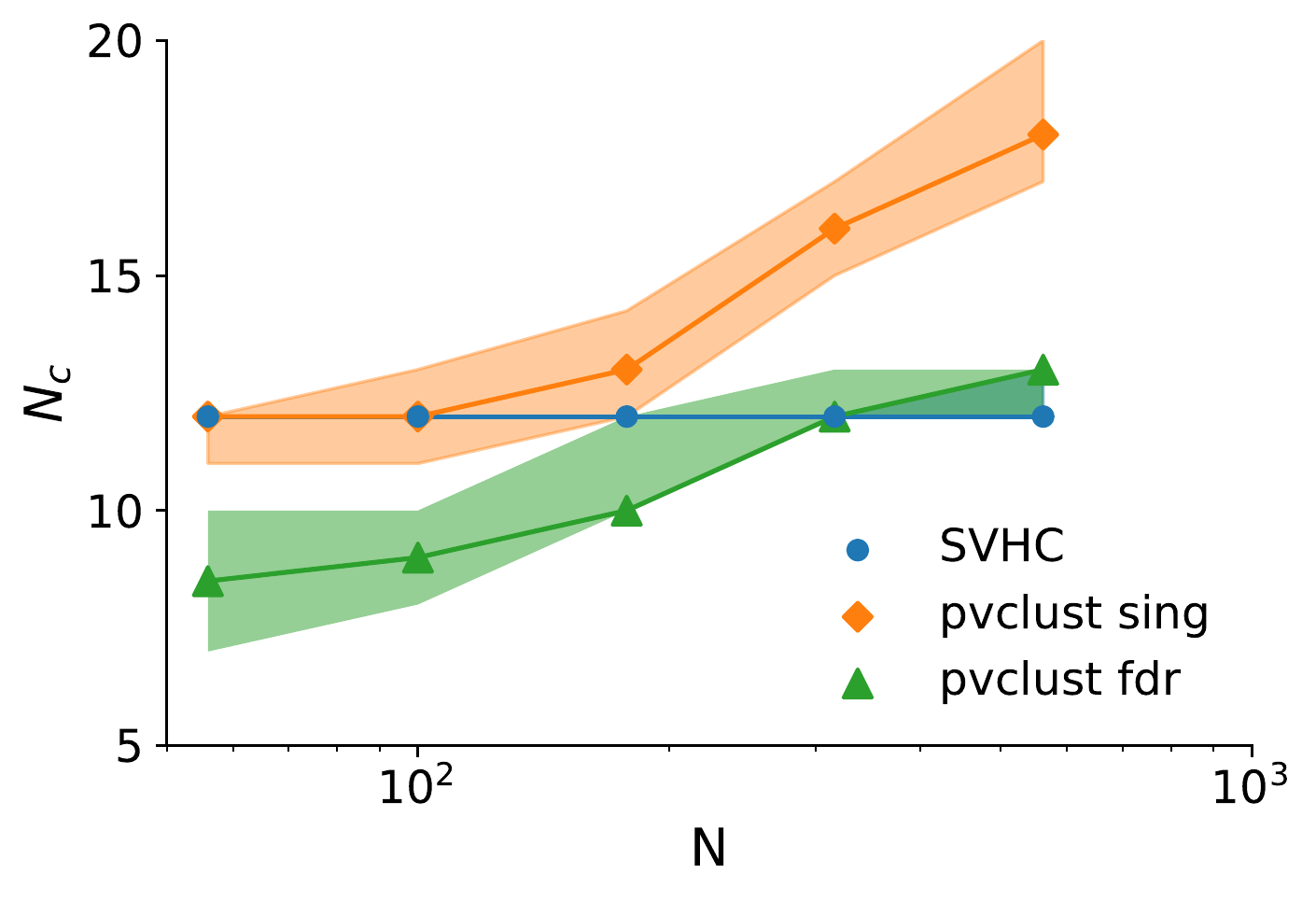} }
\subfigure[\label{fig:timeNvar}]{\includegraphics[width=.32\linewidth]{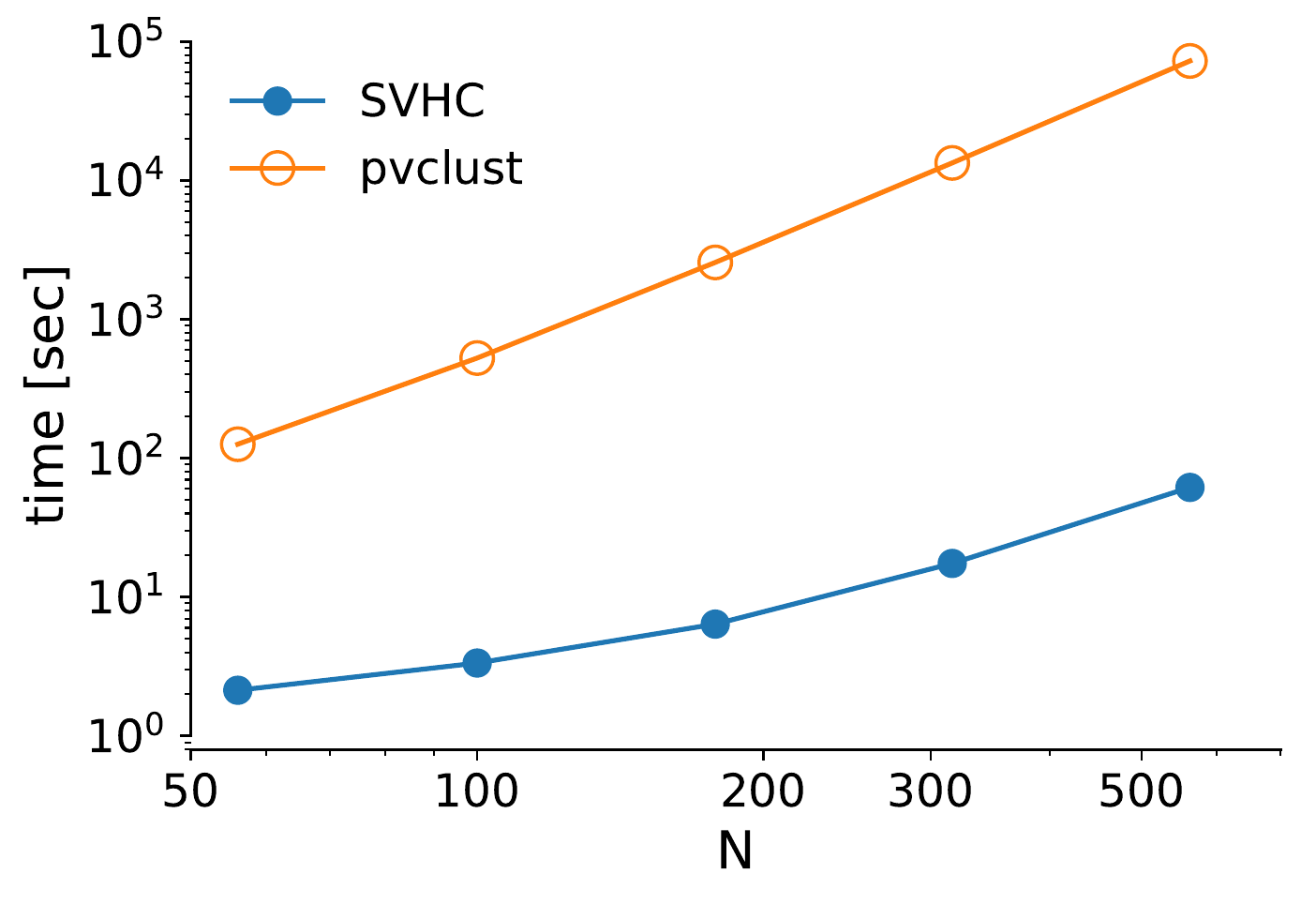} }

\subfigure[\label{fig:onmiMvar}]{\includegraphics[width=.32\linewidth]{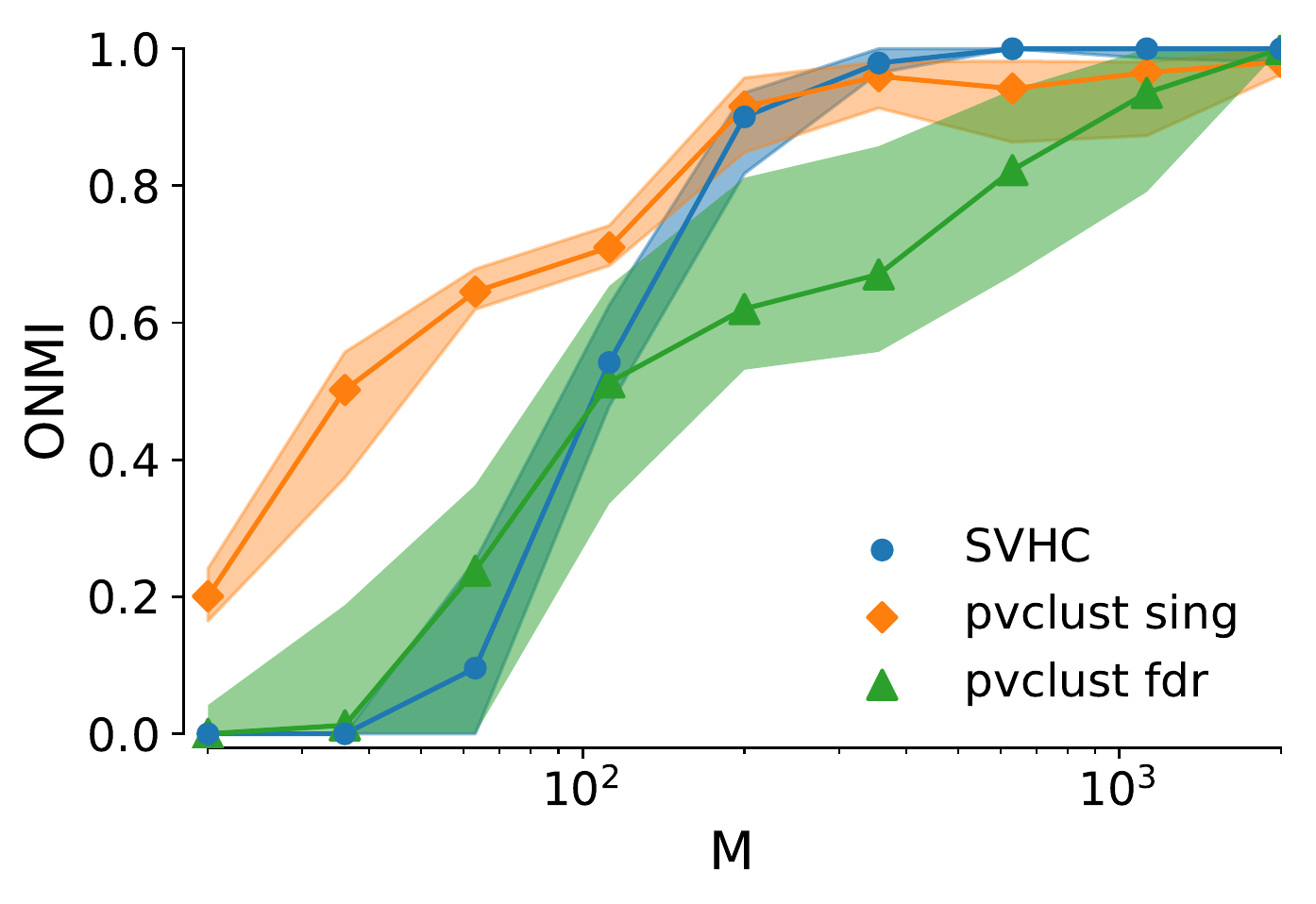} }
\subfigure[\label{fig:ncMvar}]{\includegraphics[width=.32\linewidth]{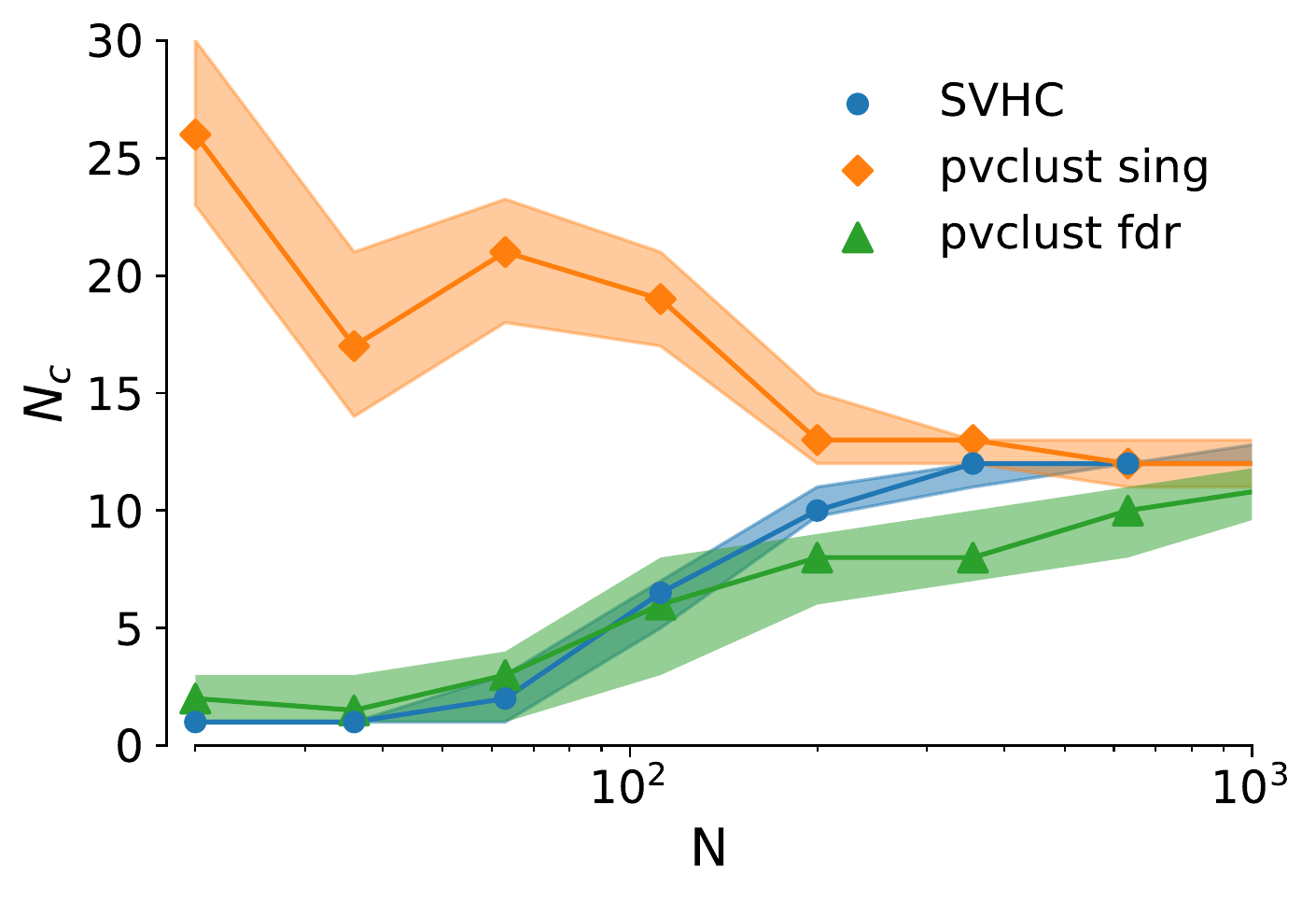} }
\subfigure[\label{fig:timeMvar}]{\includegraphics[width=.32\linewidth]{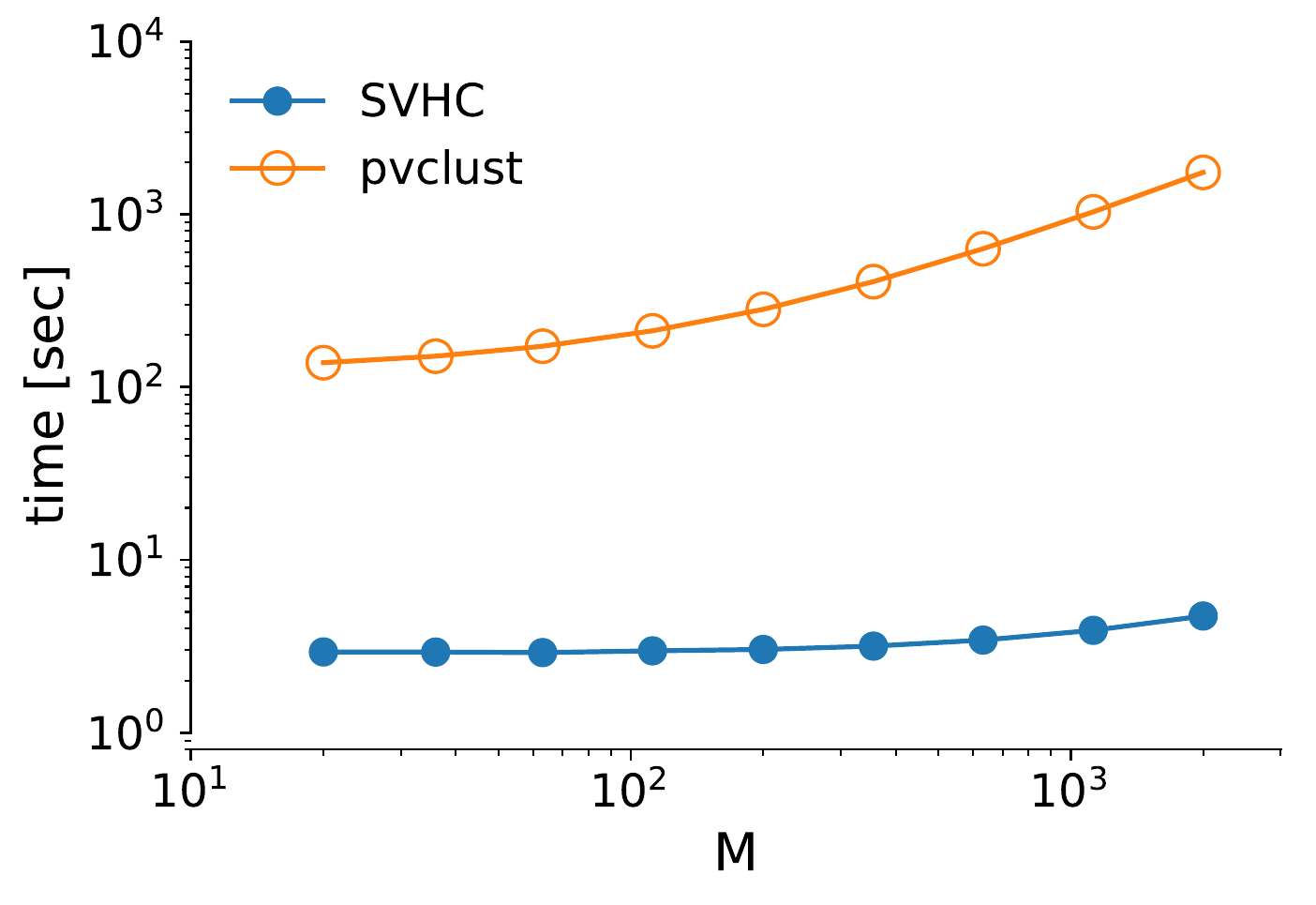} }

\caption{Numerical experiments with a benchmark of the type shown in Fig.~\ref{fig:bench} for different values of the size of the system $N$ and for different size of records $M$. $(a)$ ONMI between the true hierarchical partition of the benchmark and the hierarchical partition obtained with SVHC, Pvclust "single" or Pvclust "FDR"  as a function of the system size $N$. $(b)$ Number of statistically validated clusters detected by the algorithms  as a function of the system size $N$.  $(b)$ Computational time $T_c$ of the algorithms as a function of the system size $N$, In all simulations shown in panels $(a)$, $(b)$, and $(c)$  $M=5N$. $(d)$ ONMI between the true hierarchical partition of the benchmark and the hierarchical partition obtained with SVHC, Pvclust "single" or Pvclust "FDR"  as a function of $M$. $(e)$ Number of statistically validated clusters detected by the algorithms  as a function of $M$.
$(f)$ Computational time $T_c$ of the algorithms as a function of $M$, In all simulations shown in panels $(d)$, $(e)$, and $(f)$  $N=100$. Points are the median computed in $100$ independent realizations. The color band highlights the interval between the $25$ and the $75$ percentile. In our numerical experiments, we simulate $1000$ bootstrap replicas both for the SVHC and the Pvclust algorithm.}
\end{figure*}

Also in this case hierarchical  partitions obtained with the SVHC algorithm describes quite well the true hierarchical partition for systems with  $N$ ranging from 56 to 562 (see ONMI values in Fig.~\ref{fig:onmiNvar}).
The algorithm Pvclust has again a performance that is strongly dependent on the multiple hypothesis test correction option.  In particular, for low values of $N$ the results obtained by Pvclust "single" perform better then the results obtained with Pvclust "FDR". The reverse is true for high values of $N$. 

An analysis of the number of clusters detected by the algorithms is also highly informative (see Fig.~\ref{fig:noiseNc}) . Also for this indicator the performance of the SVHC algorithm is very good for all values of $N$. Pvclust hierarchical partitions have different characteristics for low and high values of $N$. For low values of $N$, Pvclust "single" detects a value that is very close to the true number of clusters. However, as the size $N$ increases the number of detected clusters increases too. This bias of Pvclust "single" is probably due to the absence of the multiple hypothesis test correction. In fact, the number of statistical tests performed increases linearly with the size of the system. The profile of the results obtained by Pvclust "FDR" is different. For low values of $N$ the number of clusters detected is less than the true number. This is probably due to the well known limitation of multiple hypothesis test correction. In fact the correction fully controls the amount of false positives but this is done at the expenses of not controlling the number of false negatives. For the large value of $N$ this limitation is progressively less important and the performance of the hierarchical partition of the Pvclust "FDR" algorithm becomes very good for large values of $N$. In summary, the Pvclust algorithm provides outputs recovering the true partition with one of the two multiple hypothesis test options. Specifically the "single" option works well for small systems whereas the FDR option is more appropriate for large systems.

An important aspect of the two algorithms is computational time. In Fig.~\ref{fig:timeNvar} we report the computational time for the SVHC and the Pvclust algorithms (the two multiple hypothesis test correction options of Pvclust do not significantly affect the computational time of the algorithm). From Fig.~\ref{fig:timeNvar}  it is evident that SVHC is much faster than Pvclust, and the difference in computational time increases when the size of the system increases. We numerically estimate the time dependence of computational time $T_c$ as a function of $N$ by fitting $T_C$ with a power law function $T_c = c_0 + c_1\,N^{\gamma}$ in the whole interval of $N$ values and we obtain $\gamma=1.94$ for the Pvclust algorithm and $\gamma=1.70$ for the SVHC algorithm. 
The other fitting parameters are $c_0=22$ and $c_1=3 \times 10^{-2}$ for Pvclust and $c_0=2.4$ and $c_1=3 \times 10^{-4}$ for SVHC. In addition to the difference observed in the exponent $\gamma$, it should be also noted that the coefficients $c_0$ and $c_1$ of $T_c$ for the SVHC algorithm are much smaller than the same coefficients for Pvclust.


In the last set of numerical experiments, we investigate the performance of the two algorithms as a function of the number of records $M$ of the elements of the multivariate time series. Specifically, we fix $N=100$,  $\lambda=0$ and $M = \{ 20,\, 36,\, 63, 112,\, 200,\, 356,\, 632,\, 1125,\, 2000\}$ (again a set of values with logarithmic spacing). The results summarized in Fig.\ref{fig:onmiMvar} show that SVHC outperforms Pvclust in detecting the true hierarchical partition for high values of $M$. On the contrary, for low values  of $M$ Pvclust "single" performs better than SVHC. More details about the ability of the algorithms to detect the true partition can be obtained by inspecting Fig.~\ref{fig:ncMvar}. This figure plots the number of clusters detected by the algorithms. For low values of $M$, the algorithm Pvclust "single" has a low number of false negative but this performance is obtained at the expenses of a large number of false positive, whereas both SVHC and Pvclust FDR have a large number of false negative. Depending whether the most important aspect is statistical precision or statistical accuracy the most appropriate algorithm turns out to be different. 

It is again worth noting that computational time is very different for the two algorithms and also the scalability is different. Fig.~\ref{fig:timeMvar} shows that computational time needed for SVHC and Pvclust.  We again fit the computational time with the functional form  $T_c(M) = c_0 + c_1\,M^{\gamma}$ and obtain $\gamma=0.76$ for the Pvclust algorithm and $\gamma=5.5 \times 10^{-4}$ for the SVHC algorithm. The other fitting parameters are $c_0=123$ and $c_1=1.01$ for Pvclust and $c_0=2.89$ and $c_1=1.07$ for SVHC. From the figure and from the parameters of fitting  it is quite evident that the SVHC computational time has a very limited increase as a function of $M$. In fact the fitting exponent $\gamma$ for the SVHC algorithm is very close to zero. On the contrary,  Pvclust computational time is characterized both by a sizeable exponent and also by a large minimum constant time ($c_0$). 

This set of numerical experiments confirms that SVHC algorithm is much faster and presents better scalable characteristics than Pvclust.

\subsection*{Applications to an empirical dataset}\label{sec:real}

We now apply SVHC to a widely investigated empirical dataset. As in previous numerical experiments, the application of SVHC is done in parallel with the application of Pvclust with the two options (i.e. the "single" and the "FDR" option). The dataset we investigate is a set of microarray data of lung cancer tissues. Specifically, the dataset is the gene expression pattern of $N=73$ tumor tissues belonging to 56 different patients. The data comprises information on $M=915$ selected genes. 

\subsubsection*{Adenocarcinoma of the lung data}
\begin{figure*}[tbh]
\centering
\subfigure[\label{fig:9a}]{\includegraphics[width=.32\linewidth]{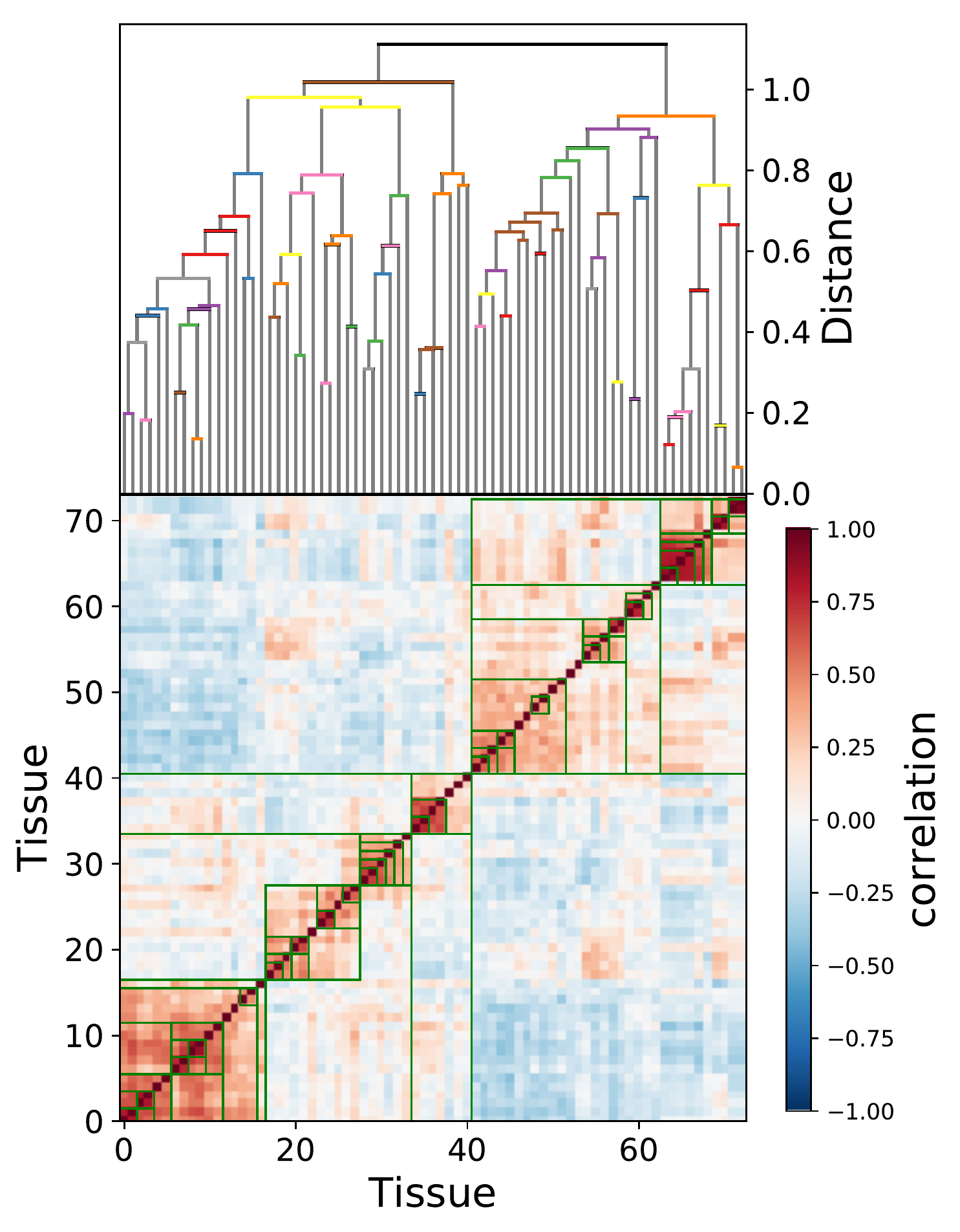}} 
\subfigure[\label{fig:9b}]{\includegraphics[width=.32\linewidth]{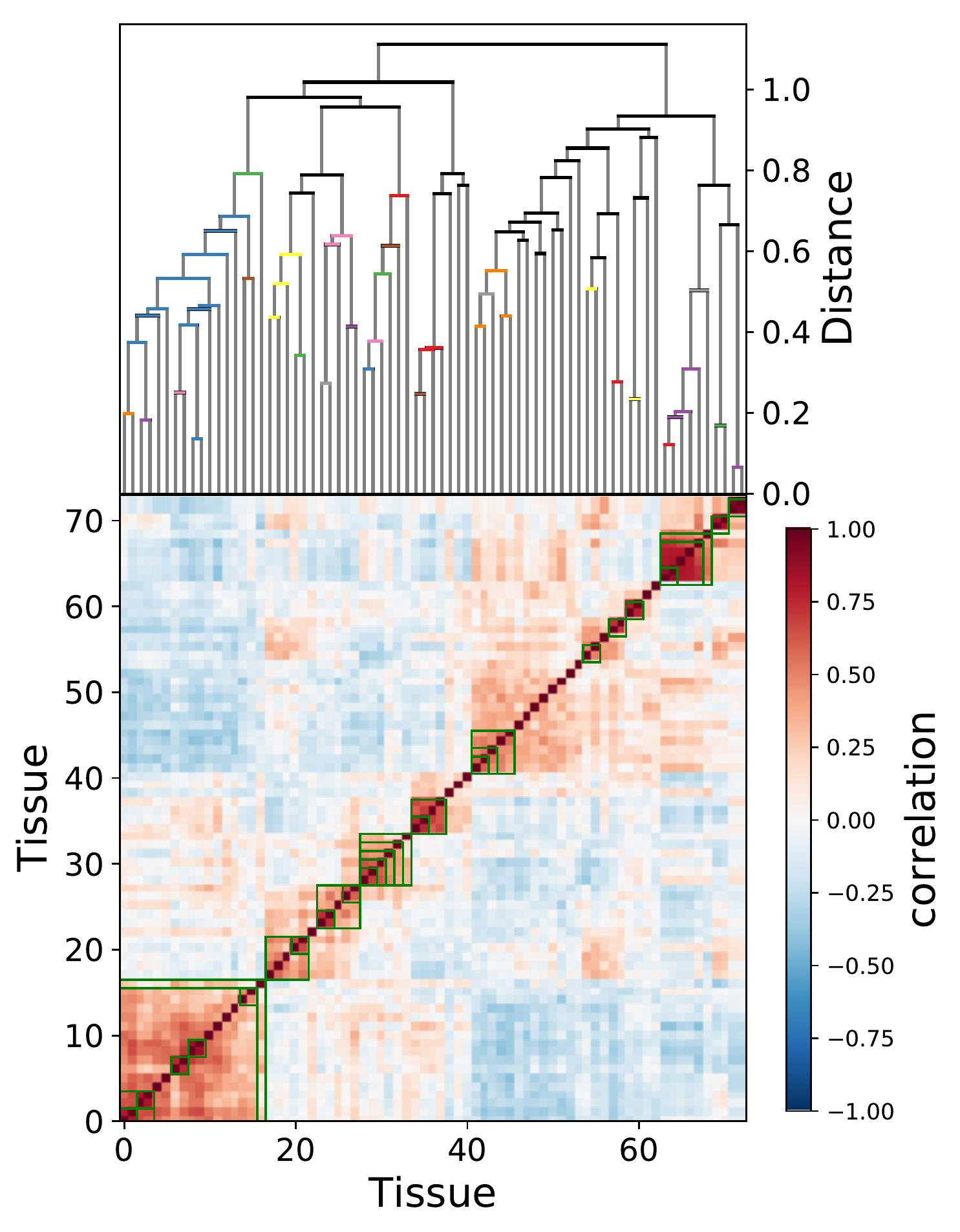}} 
\subfigure[\label{fig:9c}]{\includegraphics[width=.32\linewidth]{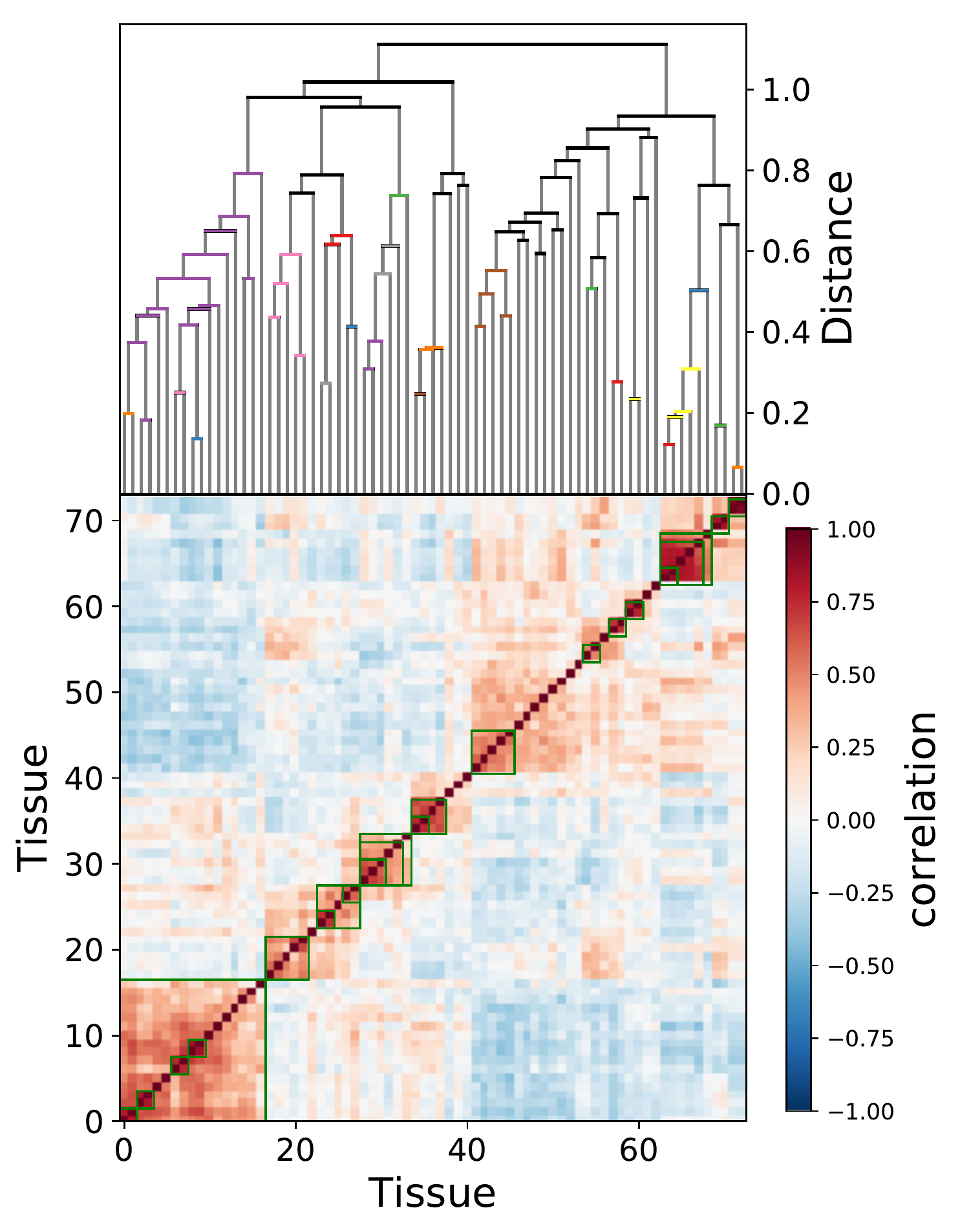}} 

\subfigure[\label{fig:9d}]{\includegraphics[width=.32\linewidth]{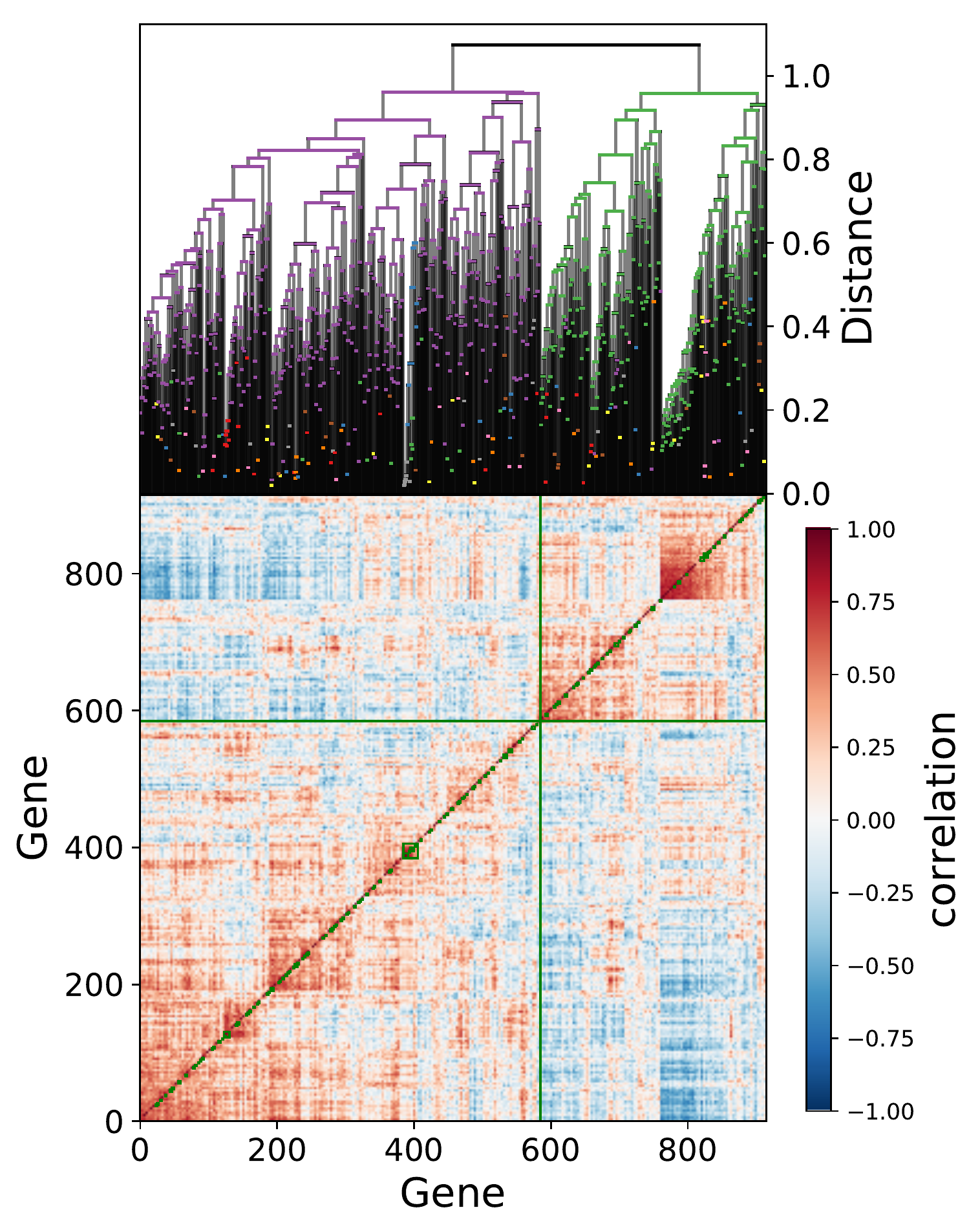}}
\subfigure[\label{fig:9e}]{\includegraphics[width=.32\linewidth]{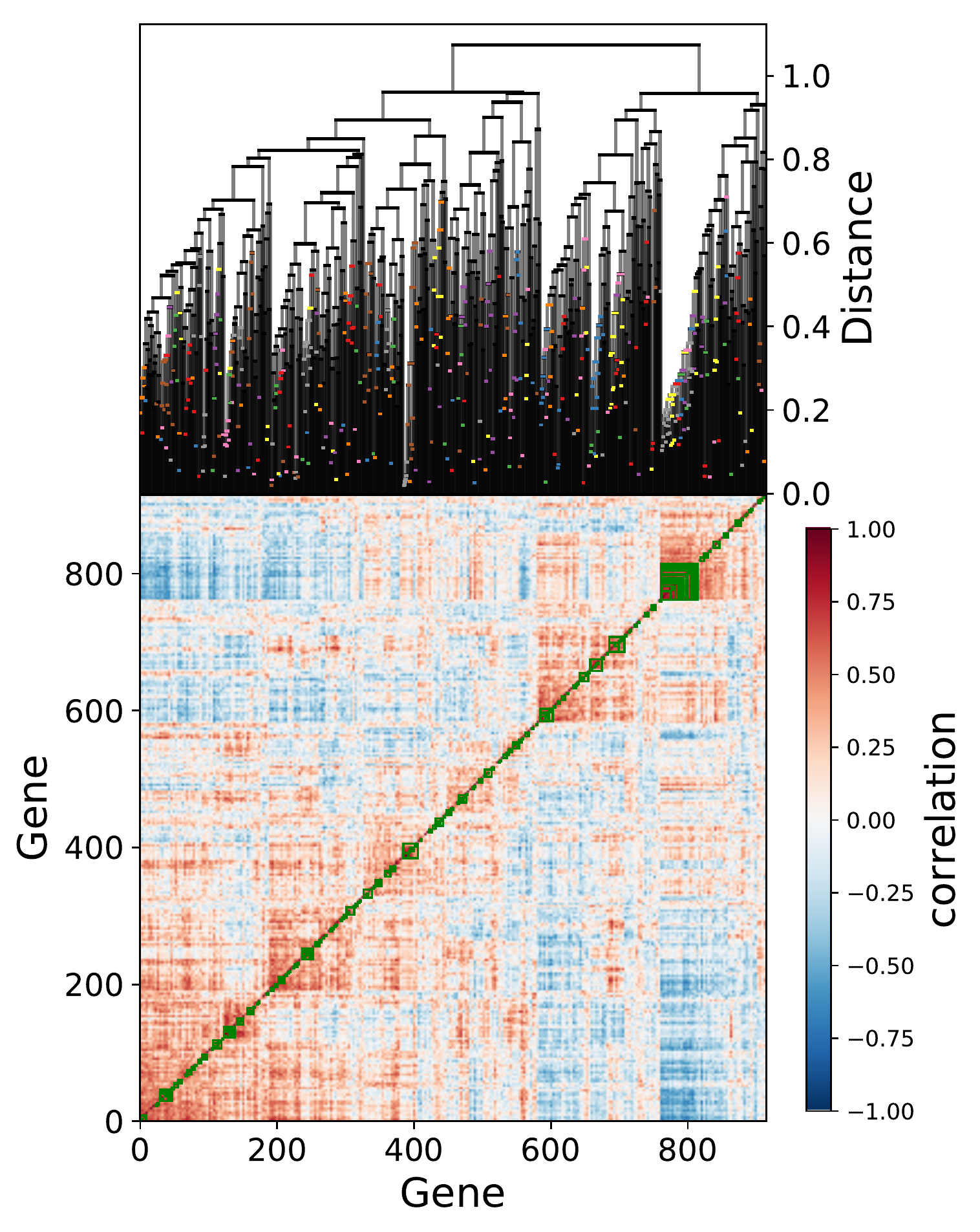}}
\subfigure[\label{fig:9f}]{\includegraphics[width=.32\linewidth]{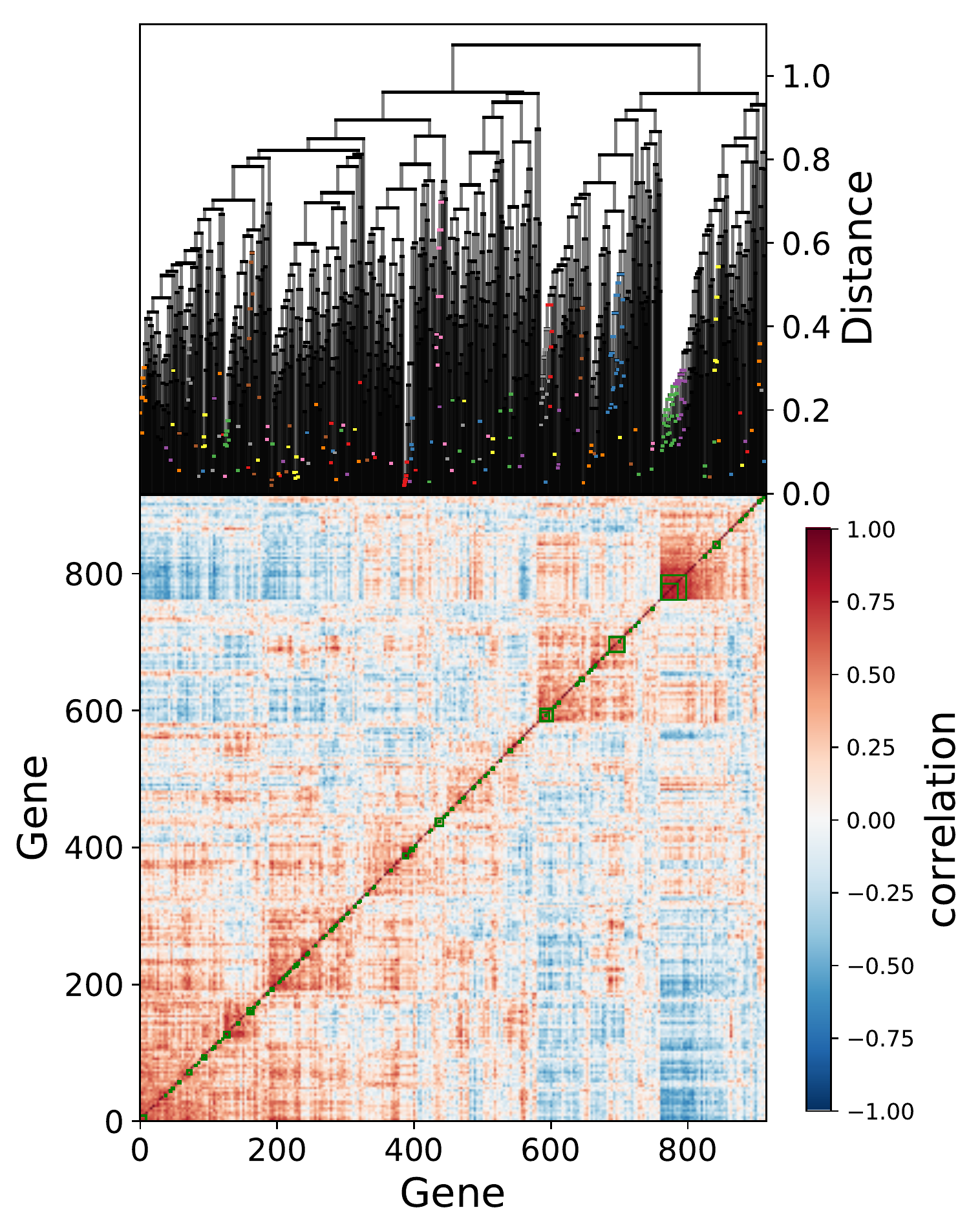}}
\caption{Hierarchical trees (average HC) and correlation matrices of lung tissues dataset (top panels) and lung genes dataset (bottom panels) \cite{garber2001diversity}. In the correlation matrices we highlight with boxes hierarchically nested clusters detected by different algorithms. $(a)$ SVHC on lung tissues, $(b)$ Pvclust "single" on lung tissues, $(c)$ Pvclust "FDR" on lung tissues. $(d)$ SVHC on lung genes, $(e)$ Pvclust "single" on lung genes, $(f)$ Pvclust "FDR" on lung genes.}\label{fig:real}
\end{figure*}

The dataset was originally collected in Ref.~\cite{garber2001diversity}, and it was used to provide an illustrative example of Pvclust performance in Ref.\cite{suzuki2006Pvclust} to obtain $p$-values of the branching points of hierarchical tree of tissues. Here we investigate both the hierarchical tree of tissues $(N=73)$ and the hierarchical tree of genes $(N=915)$ to provide both a basic example (in the case of tissues) and a more demanding example (in the case of genes) of application of the algorithms to systems of size varying more than one order of magnitude.  In our investigation, both the SVHC and the Pvclust algorithms perform $10,000$ bootstrap replicas. In Fig.~\ref{fig:real} we show the results of our investigation. A square in the matrix highlights a cluster of elements characterized by a $p$-value rejecting the statistical null hypothesis. 

To quantify the degree of similarity of partitions obtained we compute the ONMI between all pairs of hierarchical partitions obtained. For lung tissue the ONMI between the SVHC partition and the Pvclust "single" partition is ONMI(SVHC,Pvclust "S")=0.527 whereas ONMI(SVHC,Pvclust "FDR")=0.368 and ONMI(Pvclust "S",Pvclust "FDR")=0.768. 
For lung genes we obtain ONMI(SVHC,Pvclust "S")=0.321,  ONMI(SVHC,Pvclust "FDR")=0.638 and ONMI(Pvclust "S",Pvclust "FDR")=0.434. 
 By analyzing these values, we note that the hierarchical partitions obtained with SVHC are not too different from the ones obtained with Pvclust both for the small set of elements (tissues) and for the large set of elements (genes). However, in the two cases the highest degree of similarity between partitions is observed for a different option of Pvclust. Specifically, for the hierarchical partition of lung tissues the highest similarity is between the partition of SVHC and the partition of Pvclust "single" whereas for the hierarchical partition of genes the highest similarity is with Pvclust "FDR". This result again suggests that hierarchical partition of Pvclust "single" includes more false positives when the system size increases. A comparison of the hierarchical partition obtained with SVHC with the hierarchical partitions obtained with Pvclust "single" and Pvclust "FDR" can therefore provide indication about the option of Pvclust more appropriate to the size of the system investigated.    



As already noted in the investigation of the synthetic benchmark, when the system size increases  the computational time grows on. The same beahavior is observed in the analysis of empirical data. In fact, the computational time of the two algorithms for the two hierarchical trees investigated (i.e. for the lung tissues and the lung genes) is as follows: The computational time for lung tissues (i.e. a 73x915 system) is 172 s for SVHC and 4,767 s for Pvclust whereas for lung genes (i.e. a 915x73 system) is 1,434 s for SVHC and 109,660 s for Pvclust. For both investigates sets the computational time of Pvclust is much longer then the computational time of SVHC (27.7 times for lung tissues and 76.5 times for lung genes). 

\section*{Discussion}\label{sec:con}
Hierarchical clustering is a powerful data analysis tool widely used in many disciplines. The association of hard or hierarchical partitions to hierarchical trees is still an open problem. An approach widely used in genomics has its roots in phylogenetic studies. In such studies bootstrap replicas of each clade are  computed and the observation of the number of times the sample composition of the clade is detected in replicas provide a first estimation of the $p$-value to be associated to a given clade. Over the years this approach has been refined to minimize biases affecting the estimation of the $p$-value. Today the widely used Pvclust package uses this approach. Pvclust is therefore setting the standard for a statistical assessment of a specific hierarchically nested partition obtained from a given hierarchical tree. Pvclust has a great control of the statistical hypothesis underlying its approach but has also a few drawbacks. One drawback concerns the presence or absence of a procedure of multiple hypothesis test correction. From a statistical point of view, such correction should be present but the results obtained by the algorithm in the presence of a multiple hypothesis test correction are sometimes disappointing due to the fact that such a correction can be too demanding for some systems. The second and most important drawback is the computational time needed to perform the hierarchical clustering estimation for all bootstrap replicas. This time could be quite long for moderately large datasets and therefore the Pvclust algorithm is of limited use for big datasets.

In this paper we introduce a greedy algorithm that is quite effective in the detection of the true hierarchically nested partition of a multivariate series. We prove the efficacy of our algorithm by performing numerical experiments on a set of benchmarks generated by using a hierarchical factor model. The application of Pvclust to the same benchmarks show the efficacy and limitations of Pvclust with the two options of absence (i.e. Pvclust "single") and presence of multiple hypothesis test correction (i.e. Pvclust "FDR"). Our algorithm is much faster than the Pvclust algorithm and has a better scalability both in the number of elements and in the number of records of the investigated multivariate set. We therefore propose to use our algorithm in all cases when the Pvclust algorithm is too slow to be used or when it produces outputs that are quite different in the presence or absence of the multiple hypothesis test correction.

\bibliography{sample}

\section*{Acknowledgements (not compulsory)}

Acknowledgements should be brief, and should not include thanks to anonymous referees and editors, or effusive comments. Grant or contribution numbers may be acknowledged.

\section*{Author contributions statement}
C.B., S.M. and R.N.M. conceived the study. C.B. developed the algorithm and tested it on benchmarks and empirical datasets. C.B., S.M. and R.N.M. analyzed and interpreted the results obtained. C.B. and R.N.M. wrote the manuscript. All authors reviewed the manuscript. 

\section*{Additional information}

\textbf{Accession codes} A python package of the algorithm is accessible at the github web page \url{https://github.com/cbongiorno/svhc}; \textbf{Competing interests} The authors declare no competing interests.

\end{document}